# Hadamard magnetization transfers achieve dramatic sensitivity enhancements in homonuclear multidimensional NMR correlations of labile sites in proteins, polysaccharides and nucleic acids

Mihajlo Novakovic[1], Ēriks Kupče[2], Andreas Oxenfarth[3], Marcos D. Battistel[4], Darón I. Freedberg[4], Harald Schwalbe[3] and Lucio Frydman[1,*]

[1]Department of Chemical and Biological Physics, Weizmann Institute of Science, 7610001 Rehovot, Israel

[2]Bruker UK Ltd., Banner Lane, Coventry, UK

[3]Institute for Organic Chemistry and Chemical Biology, Center for Biomolecular Magnetic Resonance, Johann Wolfgang Goethe-University, D-60438 Frankfurt/Main, Germany

[4]Laboratory of Bacterial Polysaccharides, Center for Biologics Evaluation and Research, Food and Drug Administration, 10903 New Hampshire Ave, Silver Spring, Maryland, United States 20903

**Abstract**

Multidimensional EXchange, TOtal Correlation and Nuclear Overhauser Effect SpectroscopY (EXSY, TOCSY, NOESY) lie at the foundation of homonuclear magnetic resonance correlation experiments in organic and pharmaceutical chemistry, as well as in structural biology. Limited magnetization transfer efficiency is an intrinsic downside of all these methods, particularly when targeting fast-relaxing or rapidly exchanging species. These phenomena are clearly observed in systems with labile protons ubiquitous in polysaccharides, sidechains and backbones of proteins, and in nucleic acids' bases and sugars: The fast decoherence imparted on these protons through solvent exchanges, greatly reduces their involvement in homonuclear correlation experiments. We have recently discussed how these decoherences can be visualized as an Anti-Zeno Effect, that can be harnessed to enhance the efficiency of homonuclear transfers. Looped PROjected SpectroscopY (L-PROSY) resulted from these considerations, leading to ≈200-300% enhancements in NOESY and TOCSY cross-peaks for amide, hydroxyl and amine groups in biomolecules. This study demonstrates that even larger sensitivity gains per unit time –equivalent to reductions by several hundred-folds in the experiments' durations– can be achieved by looping inversion or using saturation procedures. In the ensuing correlation experiments *a priori* selected frequencies are encoded according to Hadamard recipes, and subsequently resolved along the indirect dimension via linear combinations. Magnetization-transfer (MT) processes reminiscent of those occurring in chemical-exchange saturation transfer (CEST) provide then significant enhancements in the resulting 2D homonuclear cross peaks, in only a fraction of a normal 2D experiment's acquisition time. These gains can be achieved at any fields but benefit considerably from the higher resolution and longer proton $T_1$s provided by ultrahigh field NMR. This is corroborated with experiments performed at 1 GHz, where the effectiveness of the ensuing three-way polarization transfer interplay between water, labile and non-labile protons was confirmed for proteins, homo-oligosaccharides and nucleic acids. In all cases, cross-peaks that were barely detectable in conventional 2D NMR counterparts, were measured ca. 10-fold faster *and* with 200-600% signal enhancements by the Hadamard MT counterparts. Explanations on the efficiency of these new experiments, their application to additional systems, and extensions to higher dimensionalities, are also discussed.

**Introduction**

Two-dimensional (2D) homonuclear NMR correlations[1,2] are an integral part of the tools used to elucidate the structure and dynamics of organic, pharmaceutical and biological molecules.[3,4] These correlations can be mediated by chemical exchange or by Nuclear Overhauser Enhancements (NOEs),[5,6] and are probed by monitoring how off-equilibrium polarization from one spin reservoir travels to another via dipolar interactions or chemical kinetics.[7–12,13–15] Magnetization transfers within a *J*-coupled spin network as achieved by TOtal Corelation SpectroscopY (TOCSY),[16,17] leads to complementary information based on bond connectivities. Despite being routinely performed these 2D NMR experiments, and particularly NOESY, suffer from relatively low efficiencies, leading to low signal-to-noise ratio (SNR) in their cross-peaks, and to a need for extensive signal averaging times. Detection of NOESY and TOCSY cross-peaks becomes even more difficult when involving labile sites, as information is then "lost" through chemical exchanges with the solvent. Hydroxyl protons in saccharides, amino groups in proteins and nucleic acids, amides in disordered proteins, and imino protons in RNA/DNA, are prototypical examples of such challenging systems: when placed in their natural aqueous environment all of these will undergo a rapid exchange with the solvent, that dramatically reduces the efficiency of their intramolecular polarization transfers.

We have recently introduced Looped PROjective SpectroscopY (L-PROSY),[18] an approach that alleviates these problems by regarding these exchanges as "resets" within the framework of Anti-Zeno Effects.[19–23] Instead of applying a single mixing period for homonuclear transfers that will then reach kinetically-compromised amplitudes, L-PROSY "freezes" these transfers after they begin to act with their (fastest) initial rate, resets the labile protons' states to their initial conditions by exploiting their exchange with an unperturbed solvent polarization reservoir, and repeats this process multiple times.[24,25] The ensuing 'L-PROSY encoding' acts then as a sort of conveyor belt, causing the NOE/TOCSY cross-peaks to grow with the much more favorable rates characterizing their initial buildups, and lasting for as long as either thermodynamic considerations or the recipient's $T_1$ will accommodate them –before performing the latter's signal detection. By selectively addressing only the targeted protons and avoiding water perturbation L-PROSY exploits some elements of the SOFAST NMR;[26–29] at the same time, by its repeated action, it is also reminiscent of certain CEST polarization transfer elements.[23-26] Despite their sensitivity gains, L-PROSY experiments are still long, requiring traditional $t_1$ evolution periods to build-up multidimensional information. L-PROSY acquisitions can also lead to artifacts arising from an incomplete replenishment of the targeted sites' polarization by the solvent, appearing as harmonics of genuine evolution frequencies and/or as anti-diagonal peaks. The present study demonstrates a new approach capable of alleviating both drawbacks while achieving even more complete magnetization transfer (MTs), by performing Hadamard-based[34,35] selective polarization transfers of the targeted labile protons. It is shown that, whether involving multiple selective inversions or a continuous saturation procedure, this provides the highest enhancements per unit time we have seen to NOESY, TOCSY and EXSY experiments involving labile or fast relaxing protons –ca. two orders of magnitude gains over their conventional counterparts.

**Results**

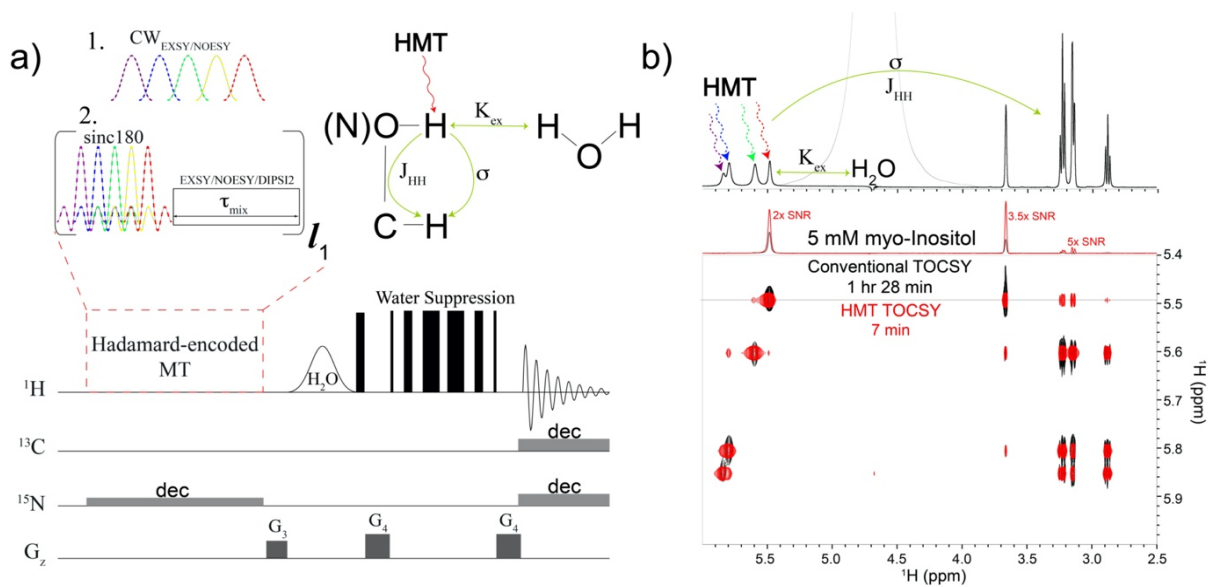

Figure 1. a) Hadamard-encoded[35] MT pulse sequence illustrating that two types of perturbation procedures can be utilized for efficient magnetization transfer according to Hadamard encoding– selective polychromatic saturations or looped polychromatic inversions (assumed here imparted by *sinc* pulses) followed by a delay for NOESY/EXSY transfer or DIPSI2[16] to achieve isotropic mixing for TOCSY (selective saturations can be used only for the NOESY/EXSY transfers; repeated inversions work for all). During these long MT processes a three-way polarization transfer is effectively established, where water constantly repolarizes labile protons enabling prolonged magnetization transfer to non-labile protons. An anti-Zeno-like effect is thus achieved. "dec" refers to GARP4[37] decoupling, that was used during Hadamard encoding and acquisition for the labeled samples. Water suppression is achieved using excitation sculpting[39] or WATERGATE 3919[40–42] schemes. b) 1D spectrum of myo-inositol showing in gray the dominating water resonance and illustrating the homonuclear transfers occurring during the MT. Shown below the 1D spectrum is a conventional TOCSY spectrum of myo-inositol acquired with 48 ms DIPSI2 mixing, overlaid on a Hadamard MT TOCSY spectrum obtained using 12 loops of 24 ms DIPSI2 mixing. Note the different acquisition times shown in the figure (≥12x faster in the HMT scheme), and the substantial (2-5x) enhancement of the cross- and diagonal peaks. Spectra were acquired at 600 MHz on a Bruker Avance III spectrometer equipped with a Prodigy probe.

*Principles of Hadamard-encoded 2D homonuclear correlations on labile sites.* Frequency-domain Hadamard spectroscopy has been proposed as a way to replace the conventional $t_1$ evolution increment of 2D time-domain NMR, by employing a "comb" of frequency-selective (polychromatic) RF pulses, that directly address peaks in the $F_1$ frequency domain.[34-36] If the frequencies of these peaks are known and well separated, 180° phase shift manipulations (Hadamard encoding) followed by suitable addition/subtraction reconstruction manipulations (Hadamard transform), have been demonstrated as means for speeding up the acquisition of 2D NMR spectra. In order to bring the advantages resulting from a replenishing solvent pool of magnetization to bear into Hadamard-encoded MT (HMT) schemes, the original manipulations were replaced by either selective polychromatic saturations or looped polychromatic inversion pulses, addressing solely the fast-exchanging, labile protons. This leads to the experimental scheme depicted in Figure 1a, where what we denote as the "HMT block" encodes via an "on" or "off" irradiation, the labile protons according to a Hadamard scheme. This is done while perturbing neither the solvent nor the peaks that will eventually receive polarization from the labile sites. The fact that the large water spin reservoir is not perturbed, provides constant repolarization of the labile protons during the encoding process. While the fact that recipients' spins are untouched prolongs the efficiency of an MT operating through cross-relaxation, *J*-coupling[38] or chemical exchange, up to times lasting on the order of the latter spins' T$_1$. Finally, water-suppressed spectra are acquired using excitation sculpting[39] or WATERGATE 3919[40–42] schemes. Figure 1b illustrates the ensuing gains, with overlaid conventional and HMT-encoded TOCSY spectra addressing the hydroxyl sites of myo-inositol –a prototypical saccharide. Whereas in conventional TOCSY the chemical exchange that labile protons undergo with the solvent averages out *J*-couplings and prevents an efficient transfer through the *J*-coupling network, in HMT the exchange does the opposite: it enhances the correlation, and magnifies cross-peaks up to 5-fold while requiring an order-of-magnitude shorter acquisition times.

To better describe and quantify HMT, the process was assessed using a Bloch-McConnell model[31,43–45] based on coupled differential equations that follow the fate of magnetization upon selective manipulation of the labile spins' pool. These labile protons were allowed to undergo suitably-population-weighted chemical exchanges with the solvent (water), while connected to a non-labile spin pool receiving polarization via a generic cross-relaxation[18] or *J*-coupling[46] process represented by a rate σ. The resulting equations can be written as:

$$\frac{dM_{y_l}}{dt} = \omega_1 M_{z_l} - (R_{2_l} + k_{ex}^l)M_{y_l} + k_{ex}^w M_{y_w}$$

$$\frac{dM_{z_l}}{dt} = -\omega_1 M_{y_l} - (R_{1_l} + k_{ex}^l + \sigma)M_{z_l} + \sigma M_{z_{nl}} + k_{ex}^w M_{z_w} + R_{1_l}M_{z_l}^0$$

$$\frac{dM_{z_{nl}}}{dt} = -(R_{1_{nl}} + \sigma)M_{z_{nl}} + \sigma M_{z_l} + R_{1_{nl}}M_{z\_nl}^0$$

$$\frac{dM_{z_w}}{dt} = -(R_{1_w} + k_{ex}^w)M_{z_w} + k_{ex}^l M_{z_l} + R_{1_w}M_{z_w}^0$$

(1)

where $M_{y_l}, M_{z_l}, M_{z_{nl}}, M_{z_w}$ are the magnetization components of the labile, non-labile and water spin reservoirs along specified axis of the Bloch sphere, and $M_{z_l}^0, M_{z\_nl}^0, M_{z_w}^0$ correspond to the equilibrium magnetizations of these reservoirs (assumed for simplicity normalized to unity). Longitudinal and transverse relaxation rates were calculated as the inverse of the corresponding relaxation times $R_{1/2} = 1/T_{1/2}$, and, in order to account for population differences between the solute and water pools, the exchange rates of the labile and water protons were scaled according to:

$$k_{ex}^l[solute] = k_{ex}^w[water] \ . \quad (2)$$

In the specific instance of a NOESY-based HMT experiment $\sigma$ represents the difference between zero- and double-quantum dipole-dipole cross-relaxation rates, and can be expressed in terms of normalized spectral densities $\mathcal{J}$ as

$$\sigma = \frac{1}{10}b^2\big(\mathcal{J}(0) - 6\mathcal{J}(2\omega^0)\big) \quad (3)$$

where $\mathcal{J}(\omega) = \frac{\tau_c}{1+\omega^2\tau_c^2}$ and $b = -\frac{\mu_0}{4\pi}\frac{\hbar\gamma^2}{r^3}$ is the dipole-dipole coupling constant.[47]

The $\omega_1$ in Eq. (1) denotes the strength of a saturation field, assumed applied along the *x*-axis. As mentioned, HMT can also be achieved using looped inversion pulses. To simulate these looped MTs within a similar model the transverse saturation terms were omitted, leaving equations that only depended on the z-components of the magnetizations:

$$\frac{dM_{z_l}}{dt} = -(R_{1_l} + k_{ex}^l + \sigma)M_{z_l} + \sigma M_{z_{nl}} + k_{ex}^w M_{z_w} + R_{1_l}M_{z_l}^0$$
$$\frac{dM_{z_{nl}}}{dt} = -(R_{1_{nl}} + \sigma)M_{z_{nl}} + \sigma M_{z_l} + R_{1_{nl}}M_{z\_nl}^0 \quad (4)$$
$$\frac{dM_{z_w}}{dt} = -(R_{1_w} + k_{ex}^w)M_{z_w} + k_{ex}^l M_{z_l} + R_{1_w}M_{z_w}^0$$

To account for the selective inversions of the labile protons an initial magnetization $M_z(0) = \begin{pmatrix} -1 \\ 1 \\ 1 \end{pmatrix}$ was taken, and numerical propagations were repeated $l_1$ times assuming that after propagating Eq. (4) for a period $\tau_{mix}$, a perturbation transformed $M_z(i \cdot \tau_{mix}) \rightarrow UM_z(i \cdot \tau_{mix})$, where $U = \begin{pmatrix} -1 & 0 & 0 \\ 0 & 1 & 0 \\ 0 & 0 & 1 \end{pmatrix}$ represents a selective labile proton inversion, and $i = 1, \ldots, l_1 - 1$.

Figure 2 shows the fate of non-labile magnetizations subject to these manipulations assuming a NOESY experiment as a case in point, and summarizes the cross-peak intensities expected from these two forms of HMT-related procedures. To better quantify the effects, the z-axes of these 3D plots are plotted as enhancements vs conventional NOESY cross-peak intensities, and calculated as $\varepsilon = M_{z_{nl}}(\tau_{sat})/M_{z_{nl}}(\tau_{mix}^{opt}, l_1 = 1)$ for a saturation-based HMT, and as $\varepsilon = M_{z_{nl}}(\tau_{mix}, l_1)/M_{z_{nl}}(\tau_{mix}^{opt}, l_1 = 1)$ for the looped inversion HMT. These enhancements are plotted vs the main parameters of these experiments: saturation time and nutation field $\omega_1$ for the continuous irradiation case (left), and number of loops and mixing time per loop for the repeated inversion experiment (column). A small molecule scenario with fast tumbling ($\tau_c = 0.1$ ns) and short internuclear distances ($r = 2$ Å) was used in the cross-relaxation rate calculations, and different chemical exchange rates $k_{ex}^l$ with the solvent were examined. In the case of a slower chemical exchange the expected enhancements reach factors ≤5, while for faster exchange rates HMT boosts cross-peak intensities by an order of magnitude vs conventional NOESY. This is understandable since, as long as they take place in the slow- to mid-rate exchange regime (i.e., as long as solvent and solute lines are well separated), faster chemical exchanges will detract from conventional cross-relaxation transfers but enhance HMT's efficiency by providing a more rapid and complete repolarization of the labile site from the abundant water pool. Chemical exchange rates also affect other aspects of the HMT enhancement: For slow $k_{ex}^l$ even low $\omega_1$ fields suffice to provide efficient saturation of the labile protons and high enhancements (Fig. 2a), while the enhancements need and benefit from higher $\omega_1$s when the repolarization by chemical exchange becomes fast (Fig. 2c). Furthermore, when relying on looped inversions, faster chemical exchanges repolarize labile protons more quickly, shifting the maximal transfer enhancements towards shorter mixing times and higher numbers of loops. Additional features including dependencies of HMT's enhancements on correlation times, internuclear distances and chemical exchange rates are summarized in Supporting Figures S1 and S2. Predictions of a similar model focusing on the enhancements anticipated for J-based HMT TOCSY transfers, are presented in Supporting Figure S3.

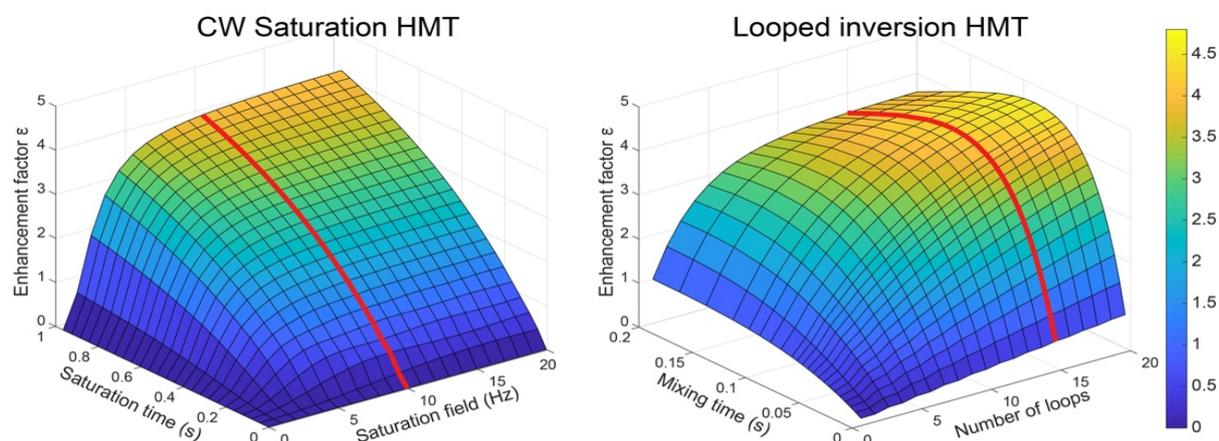

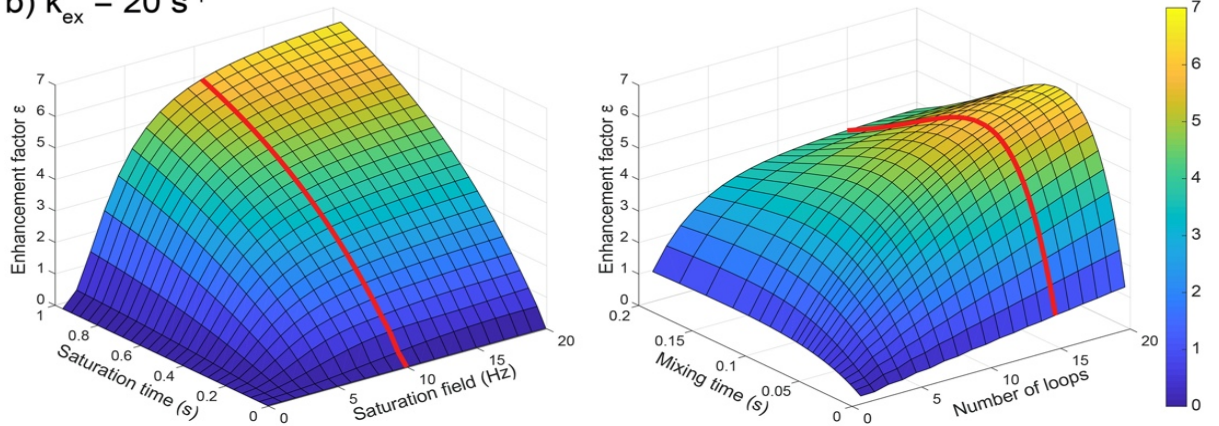

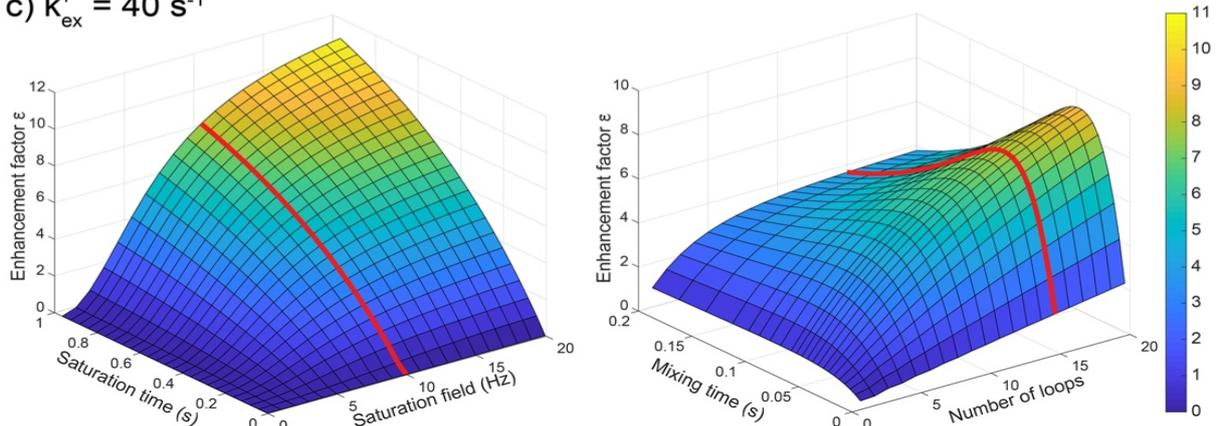

Figure 2. Enhancement factors $\varepsilon$ calculated for CW saturation (left) and looped inversion (right) versions of HMT-based experiments vs conventional NOESY transfers, computed for different rates of chemical exchange with the solvent. Cross-relaxation rates were calculated at 14.1 T for $\tau_c = 0.1\,ns$ correlation times and $r = 2\,\text{Å}$ internuclear distance. Labile and non-labile sites relaxation rates were chosen as $T_{2_l} = T_{2_{nl}} = 0.3\,s$, $T_{1_l} = 0.5\,s$ and $T_{1_{nl}} = 0.8\,s$; water relaxation constants were taken $T_{2_w} = 0.5\,s$ and $T_{1_w} = 3\,s$. A water excess of 500-fold was assumed. Red curves illustrate the conditions that in terms of saturation fields and number of loops, were usually explored in this study's experiments.

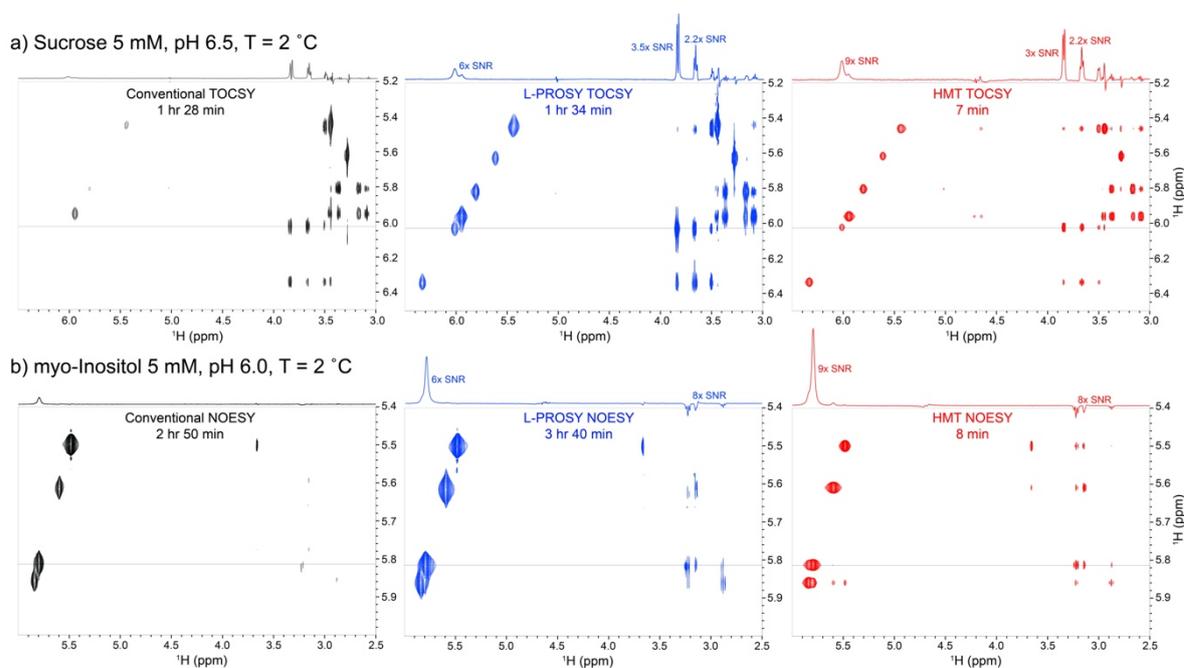

Figure 3. a) TOCSY spectra acquired on 5 mM sucrose using conventional (48 ms DIPSI2), L-PROSY (11x20 ms DIPSI2) and Hadamard MT TOCSY (12x20 ms DIPSI2) schemes. Shown on top of each 2D spectrum are 1D traces extracted at the indicated horizonal dotted lines, with numbers indicating SNR improvements vs the conventional spectrum. Notice the superior spectral quality in both the L-PROSY and MT Hadamard experiments vis-à-vis the conventional TOCSY acquisition. b) Similar comparison but for NOESY experiments on 5 mM myo-Inositol. The conventional NOESY experiment was acquired with 80 ms mixing, L-PROSY used 14 loops, 35 ms each; Hadamard MT was acquired using 800 ms CW saturation. All data were acquired on a 600 MHz Avance III Bruker equipped with Prodigy probe.

*HMT vs L-PROSY vs Conventional 2D TOCSY/NOESY.* HMT's principles are similar to those underlying L-PROSY –with main differences arising from their data acquisition formats, rather than in their cross-peak forming principles. It is therefore relevant to include the L-PROSY experiment in comparisons between HMT and conventional acquisitions. Figure 3a illustrates 2D TOCSY NMR collected with all these methods on a 5 mM sucrose sample at 2 ºC. L-PROSY yields ≈2-4x sensitivity enhancements compared to conventional experiments; HMT yields similar enhancements for every cross peak. Figure 3b shows another comparison for a NOESY experiment acquired on 5 mM myo-inositol; again, notice the ≈8-fold enhancements arising upon comparing L-PROSY and HMT with conventional NOESY, and the resemblance between the first of these two sets in terms of enhanced diagonal- and cross-peak SNRs. In both cases the Hadamard encoding required acquisitions that were shorter by an order of magnitude; this multiplexing advantage is as in previous forms of 2D Hadamard spectroscopy[35,36] and, as in the latter, it is applicable in cases where peaks are sufficiently resolved and their nature *a priori* known.

*HMT and the effects of increased magnetic field strengths.* Like CEST,[48–50] HMT-based methods that target labile protons could benefit significantly from operating at the highest possible magnetic fields. Under these conditions (i) more rapid exchange rates can be accommodated, leading to more complete replenishments from the solvent reservoir without resolution penalties (in ppm); (ii) T$_1$ relaxation times tend to get longer in, facilitating the extent of the intramolecular transfers;[47,51,52] (iii) resolution between sites improves, leading to more facile conditions for implementing the 1D Hadamard encoding and to the possibility of relying on more intense $\omega_1$ saturating fields; and (iv) it becomes generally easier to study the exchanging protons closer to physiologically-relevant temperatures. These advantages reinforce one another when studying polysaccharides, as depicted in Figures 4a and 4b for the hydroxyl protons of α2,8-linked sialic acid tetramer ((SiA)$_4$). Notice how field improves the 1D spectral resolution for these sites leading, at 5 °C and 1 GHz, to a nearly full resolution of all hydroxyl protons in the spectrum. Figures 4c and 4d compare conventional and HMT NOESY spectra acquired on the (SiA)$_4$ glycan under such conditions, when recorded at two different fields. It is hard to discern correlations involving the hydroxyl protons in the conventional experiments due to weak cross-peaks and pronounced $t_1$-noise: at 600 MHz barely any cross-peaks with these -OHs show up. Even at 1 GHz, the fast chemical exchange only permit conventional NOESY correlations between hydroxyl protons and some of the nearest aliphatic neighbors positioned at internuclear distances ≤2 Å, like IVOH8-IV8 (at 1.7 Å) and IVOH8-IV6 (at 1.9 Å). By contrast, HMT NOESY at 1 GHz –and, despite the challenges of selective saturation at lower fields, even at 600 MHz – provides a wealth of quality data, and in a fraction of the time needed by its conventional counterpart. Based on a previously solved structure of (SiA)$_4$ under super-cooled conditions,[53] these new data can be used for both assignments and for structural refinements –the former including intra-residue correlations, but the latter non-nearest neighbors. HMT reveals in fact *inter-residue* cross-peaks that are close to 5 Å apart, including IIOH7-I8 cross-peaks at ≈4.7 Å, IIOH9-I8 at ≈5.0 Å, and other cross peaks labeled red in Fig. 4d. Notice moreover, that these correlations involve hydroxyls with a wide rate of solvent exchange rates –ranging from ~10 s$^{-1}$ for IVOH8, to ~40 s$^{-1}$ for IIOH7 and even to ~100 s$^{-1}$ for IIOH9.[54] The reason for this efficiency is as explained above: faster exchanges may hurt cross-relaxation, but supply fresh polarization for the repeated transfer of information (Figs. S1, S2). Further examples presenting similar advantages for this compound but focused on TOCSY correlations, are presented in the Supporting Information (Fig. S4).

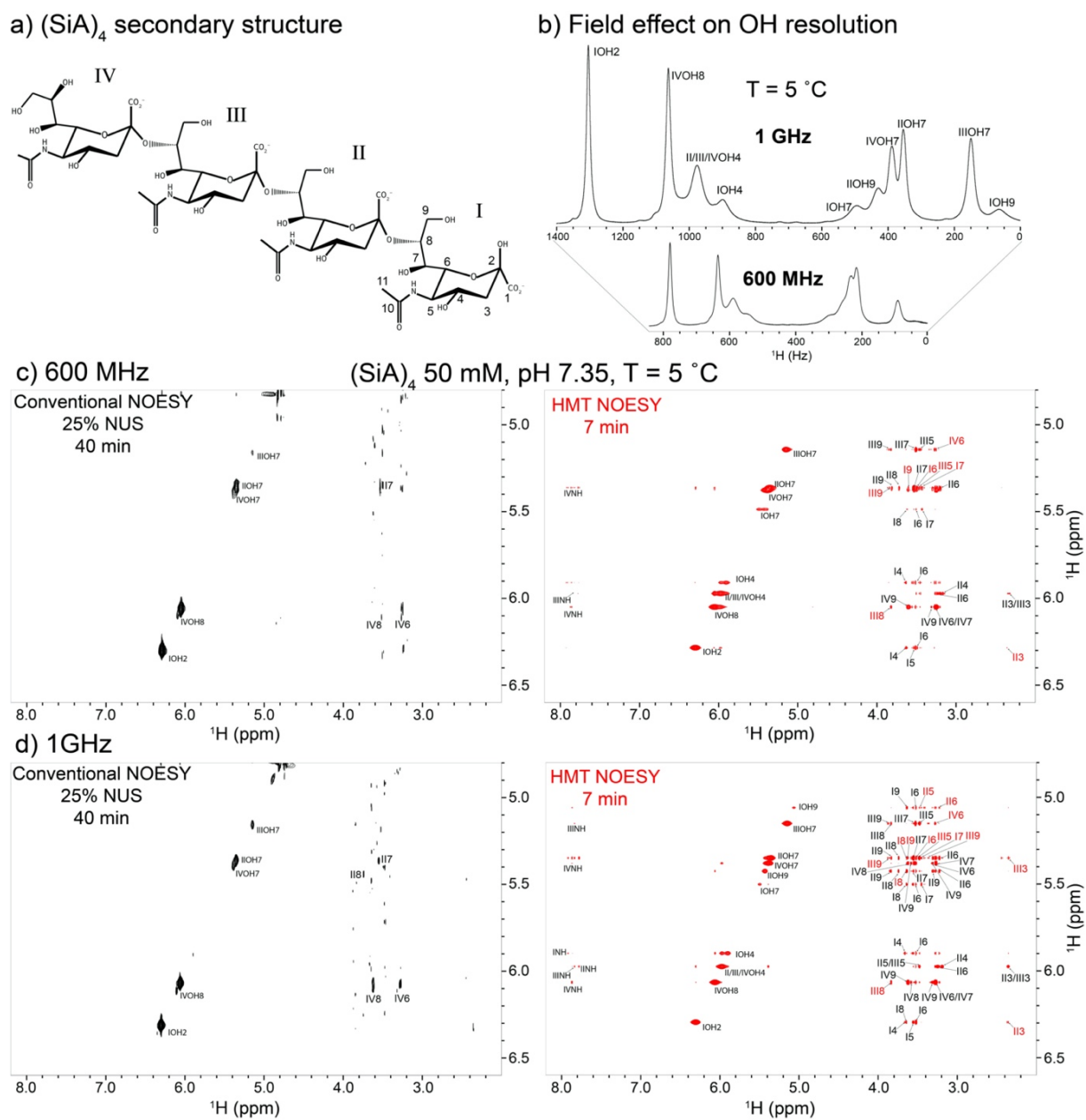

Figure 4. a) Primary structure of α2,8-linked sialic acid tetramer, (SiA)$_4$, depicting the rings' numbering and the multiple hydroxyl and amide labile $^1$Hs. b) Magnetic field effect on the linewidth of hydroxyl hydrogen atoms. Note how the chemical exchange broadens certain hydroxyl protons at 600 MHz, while at 1 GHz most of the peaks are well-resolved, thanks to field-induced separation of the exchanging sites. (c,d) Comparisons between conventional 2D NOESYs acquired using 25% non-uniform sampling for a minimal acquisition time with faithful spectral reconstruction, and Hadamard MT experiments collected with 600 ms saturation at 600 MHz and 1 GHz, respectively. While conventional NOESY only revealed the closest neighbors, much richer information (coupled to much shorter acquisitions) is provided by the HMT. Still, full unambiguous assignment of this homopolymer is only possible with the resolution arising at 1 GHz. 600 MHz spectra were acquired using an Avance III Bruker with a Prodigy probe, and at 1 GHz on an Avance Neo equipped with a TCI cryoprobe

*Amide, amino and imino proton correlations in proteins and nucleic acids.* HMT at ultrahigh magnetic field turns out to be especially informative when implemented on the imino protons of nucleic acids. At 1 GHz these imino resonances, which can be broadened by chemical exchange with the solvent at lower fields, tend to be sharp and fully resolved. Figure 5 shows the superiority of the HMT experiment over conventional NOESY for detecting cross-peaks involving these imino resonances, utilizing a 14mer hairpin RNA as prototypical example. The wealth of peaks in the HMT correlations opens the possibility of elucidating their origin: with them, it is possible to obtain nearly full spectral assignments, confirming many more correlations than what possible with conventional experiments. Particularly valuable are the correlations involving U7 and U8 in the loop region: HMT enabled assignment of these imino resonances, by providing cross-peaks that fast exchanges with the solvent made undetectable in the traditional experiment. Comparisons of conventional and HMT NOESY experiments collected for this 14mer at two different magnetic fields, are further illustrated in Figure S5. Slices extracted in these data for various imino sites confirm sensitivity/unit-time gains for the ensuing cross-peaks ranging between 200 and 600-fold. Notice moreover, that besides a superior resolution, the enhancements provided by HMT vs the conventional experiments are larger at 1 GHz that at 600 MHz. This can be attributed to one of the aforementioned field- derived advantages: at 1 GHz, the inversion/saturation pulses addressing the imino can be made more effective, leading to a more efficient MT perturbation and to ~25% larger enhancements when assessed against the conventional baseline.

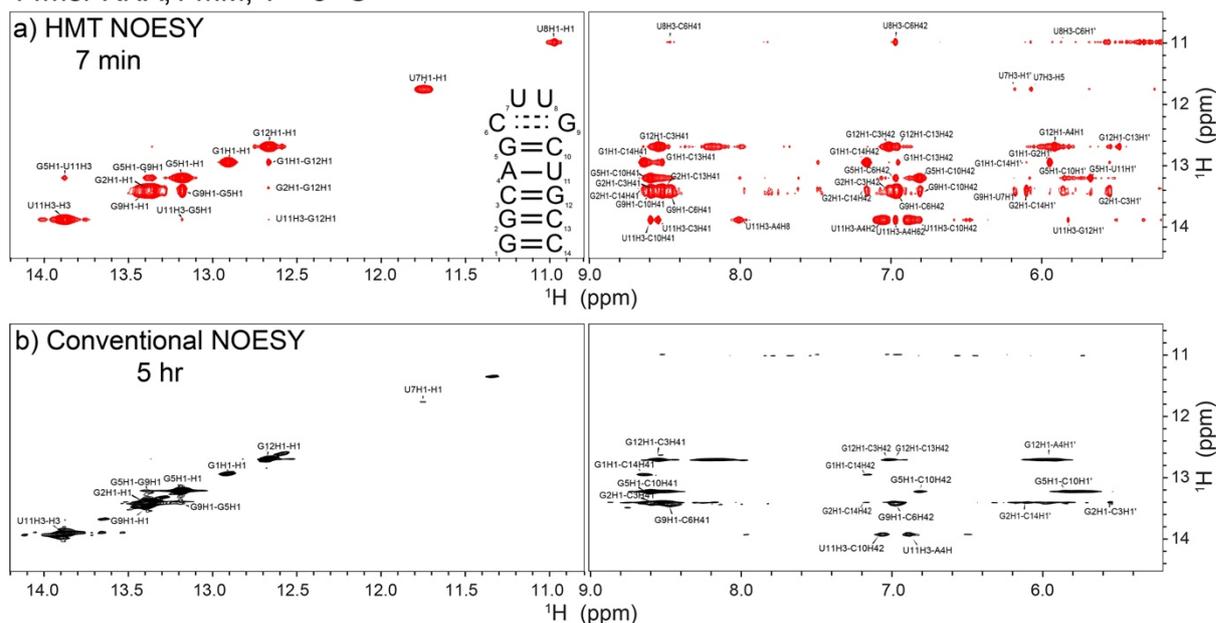

Figure 5. a) Assigned conventional (100 ms mixing, 30 us delay for a binomial water suppression optimized for imino resonances) vs. b) HMT NOESY spectra (14 loops with a 20 Hz inversion pulse followed by 40 ms mixing each). Both spectra show correlations against imino resonances that were placed along the vertical direction; for maximizing its efficiency, however, the conventional spectrum was acquired with the imino resonances along the direct domain, and was flipped in this graph to match the HMT data (that actually encoded the imino protons in F1) –otherwise, the conventional experiment showed no peaks originating from the imino altogether. As this was an [15]N-labeled sample, looped inversions / mixing periods were preferred over continuous saturations; otherwise, the constant [15]N decoupling during the long saturation period led to sample heating. Notice the large number of imino correlations with other imino, amino and sugar aliphatic protons in the HMT experiment that are undetectable in the conventional counterpart. Assignments are labeled in the spectra and corresponding structure is shown in the inset; further examples of the SNR/unit_time arising in this RNA system are shown in the Supporting Information. Spectra were acquired at 1 GHz using a Bruker Avance Neo spectrometer equipped with a TCI cryoprobe

Another advantage of operating at ultrahigh fields is that most amide backbone signals in small and medium-sized proteins become well resolved in 1D experiments, enabling HMT experiments on these N-bound [1]H resonances as well. Figure 6 illustrates the benefits resulting from this with comparisons between conventional and HMT NOESY experiments recorded on LA5, a 40-residue protein,[55] and on ubiquitin, a 76-residue protein. As in the case of the RNA these HMT experiments utilized a polychromatic looped encoding instead of single long saturation pulses, as both of these proteins were [15]N labeled and looping facilitated heteronuclear decoupling during the encoding (sequence in Fig. 1a). Extracted 1D projections show a 1:1 match between the HMT and conventional spectra, with sensitivity enhancements of ≈2x provided by the faster former scheme. These enhancements are smaller than those observed for saccharides and nucleic acids due to the slower exchange rates that amide protons exhibit in these proteins' structured environments; still when combined with 5-fold shorter acquisitions, it is clear that HMT also brings substantial sensitivity gains per-unit-time to proteins. Even larger enhancements could result on disordered proteins; in such cases, however, limited spectral dispersion would require the incorporation of a 3rd, heteronuclear-encoding dimension in order to resolve the amide peaks. Such experiments will be considered in a separate study.

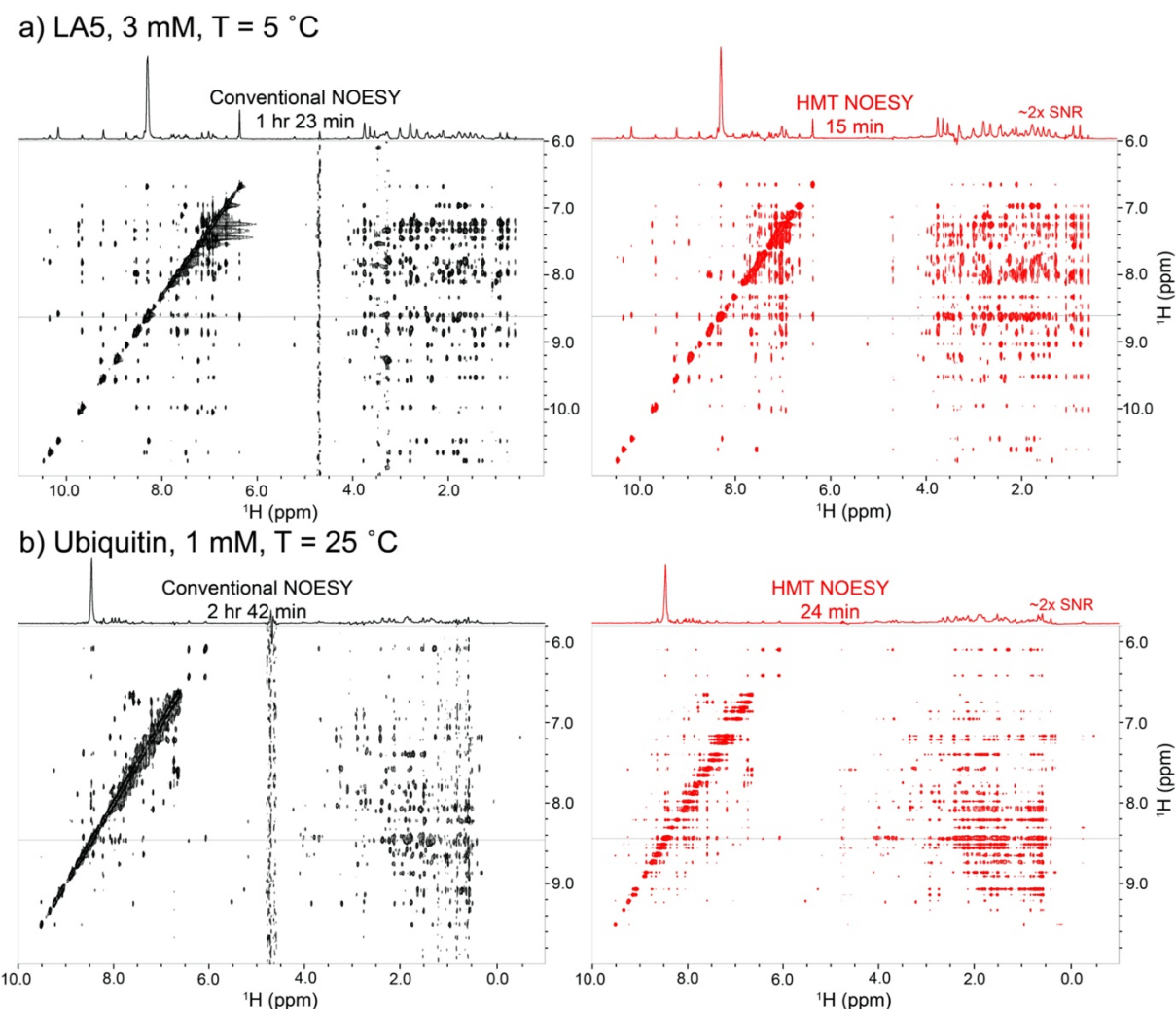

Figure 6. Conventional vs HMT NOESY spectra recorded for a) LA5 (top), and b) Ubiquitin protein samples. Notice how the 23.5 T field is sufficient to resolve almost entirely the amide/amino resonances in these structured peptides (regions between 6.6 – 9.5 ppm) and enable fast, highly sensitive NOESY experiments by Hadamard MT. Conventional experiments were acquired with 300 ms and 250 ms mixing, respectively (for maximum NOESY cross-peaks); HMT employed 6x150 ms and 6x140 ms looped encoding. Spectra were acquired at 1 GHz using a Bruker Avance Neo console equipped with a TCI cryoprobe.

The strong correlations that HMT can deliver also open up the possibility of exploiting homonuclear correlations against amino groups, whose protons are sometimes underutilized in NOESY and TOCSY experiments in both proteins and nucleic acids. Figure 7 shows how such 2D correlations can be put to good use, with HMT NOESY examples targeting amino groups in both ubiquitin and in the 14mer RNA sample. Strong correlations are observed in both spectra among the amino protons themselves, arising from combined chemical exchange and Overhauser effects among these moieties. Similar correlations are detected, much more weakly, in the conventional experiments depicted in Fig S6c for ubiquitin and in Fig S7b for RNA. Moreover, HMT experiments reveal many additional long-range NOE-driven cross-peaks with the aliphatic protons for the protein case, and with the imino protons of nucleotides that are both in the same and in neighboring base pairs throughout the nucleic acid chain. When compared to the amide protons, it is clear that the faster chemical exchanges of these amine sites endow their HMT data with larger sensitivity gains –again, in the several hundred-fold, if considering SNR / unit time.

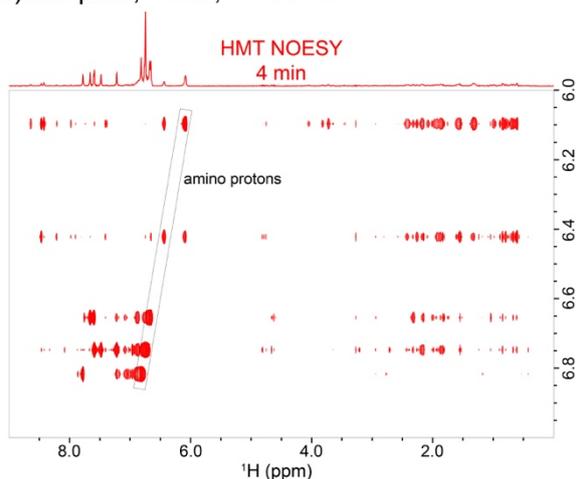 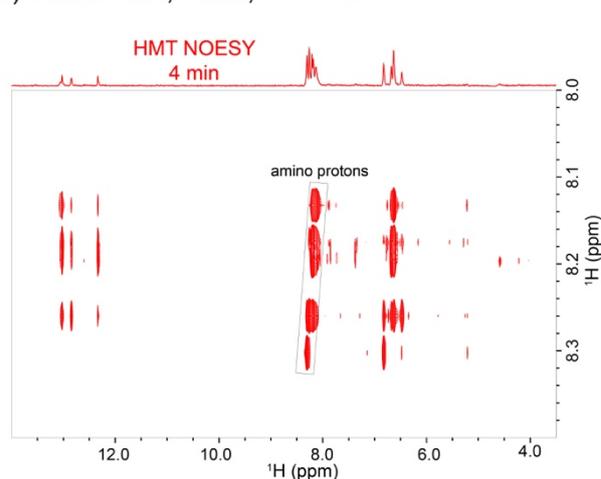

Figure 7. Amino protons targeted with Hadamard MT NOESY experiment in a) Ubiquitin and b) the 14mer RNA in Figure 5. The spectrum in a) was acquired with 12 loops and 80 ms per loop, while in b) a 600 ms long saturation pulse was used instead. The strongest correlations correspond to exchange and NOE cross-peaks to aliphatic and amino protons respectively; notice in the protein and the RNA, however, the interesting amide and imino cross-peaks emerging as well. Spectra were acquired at 1 GHz using a Bruker Avance Neo console equipped with a TCI cryoprobe.

*HMT: Exploiting hydroxyl correlations in proteins and nucleic acids.* In addition to nitrogen-bound labile protons, hydroxyl protons in sidechains and in sugars are notoriously challenging targets to work with in protein and nucleic acid NMR, respectively.[56–58] In both cases, the -OH peaks are often buried under other, sharper and more intense amine and amide resonances. On the other hand, hydroxyl protons usually undergo faster chemical exchanges with water than the latter; as illustrated in Figures 1-4, this qualifies them for potentially large cross-peak enhancements when targeted by the HMT scheme. Figure 8a shows a version of the HMT pulse sequence that could allow such usually 'hidden' hydroxyl hydrogens be encoded, by incorporating additional $^{15}$N/$^{13}$C-based filters aimed at suppressing the intense signals from protons bound to $^{15}$N and $^{13}$C that would otherwise complicate the OH's observation. To investigate if the OH protons could be targeted in such experiments, a series of 1D variable-temperature (5-25 ºC) $^{15}$N/$^{13}$C-suppressed spectra were acquired; proton resonances that survived the N-H and C-H suppression and became sharper at lower temperatures due to slower chemical exchanges with water, and were confirmed as likely candidates to arise from the labile hydroxyl protons (Fig S6b). Figures 8b and 8c exemplify the cross-correlations arising from this experiment on a doubly $^{13}$C/$^{15}$N-labeled ubiquitin and on a 14mer RNA sample, respectively. Highlighted in these spectra are the hydroxyl hydrogen atoms that are addressed in the HMT NOESY; particularly interesting are OH-OH inter-residue correlations detected in ubiquitin, and the long-range correlations between the OHs of sugar OH and the base protons resonating between 7 and 8 ppm in the 14mer RNA that are mostly missing in the conventional spectrum in Fig S7c. These include interesting correlations between the 2'-OH groups and the aromatic protons of the nucleotide in the 5'-direction.[56,59] In the conventional NOESY only a

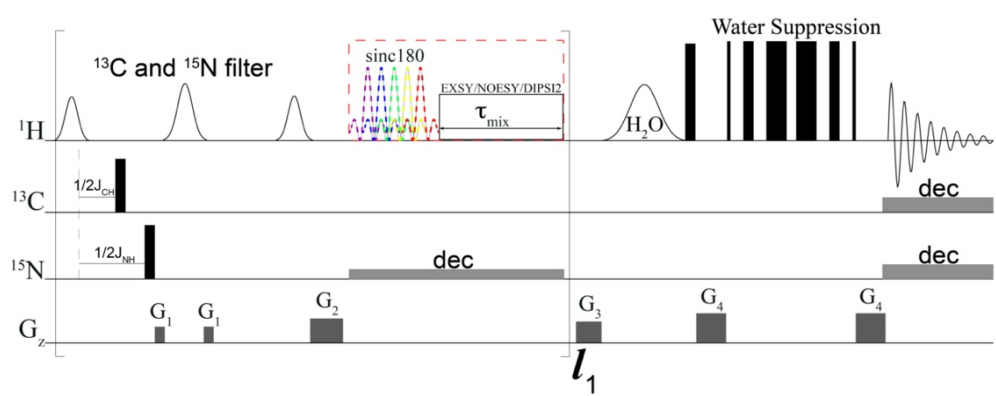

a) $^{13}C/^{15}N$ filtered HMT

b) Ubiquitin, $^{15}N/^{13}C$, 1 mM, T = 5 °C

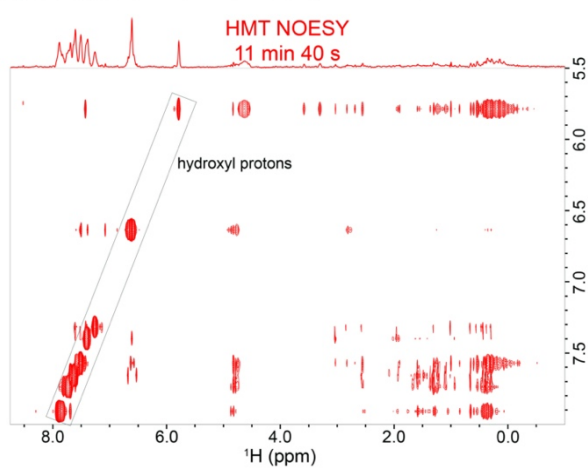

c) 14mer RNA, $^{15}N/^{13}C$, 1 mM, T = 5 °C

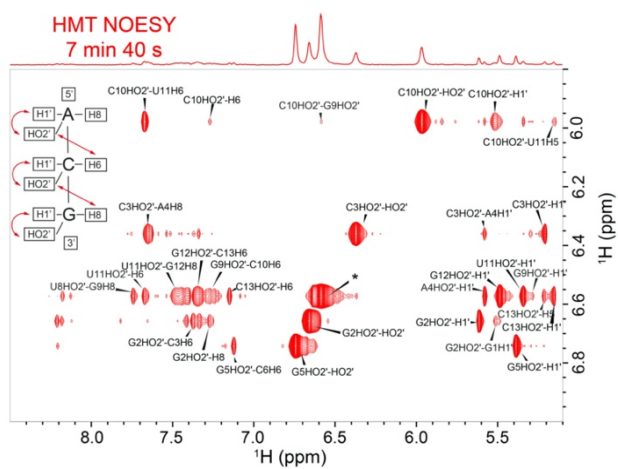

Figure 8. a) Hadamard MT pulse sequence incorporating a filter to suppress protons bound to $^{13}C$ and/or $^{15}N$ in labeled compounds, in order to selectively observe 'hidden' hydroxyl hydrogens. The filter is applied at the beginning of every MT Hadamard loop in order to prevent the suppressed $^{13}C$- and/or $^{15}N$-bound protons to start recovering during mixing periods. Consequently, short mixing times have to be used per loop, which is not a hinder in this case thanks to the OH's fast chemical exchanges with water. In the illustrated filter, a selective proton spin-echo is applied only to the HO-bearing regions, so as not to disturb other protons that maybe receiving the magnetization from these targeted sites. Examples are illustrated with b) Ubiquitin acquired using 10 loops and 50 ms per loop, and c) the 14mer RNA using 15 loops and 40 ms mixing time per loop. The structure in the inset illustrates the latter's expected NOE correlations involving hydroxyl protons, and a nearly complete assignment of these resonances revealing the many long-range correlations provided by the HMT spectrum. Resonances labeled with an asterisk does not have unambiguous assignment, as they involve 4 or 5 overlapping hydroxyl protons. Spectra were acquired at 1 GHz using a Bruker Avance Neo console equipped with a TCI cryoprobe.

single such interaction is observed, between the 2'-OH group of C13 and the aromatic proton of C14. By contrast, in the HMT NOESY, a total of *twelve* such correlations were detected, giving hitherto unprecedented inter-residue correlations and distance information –including *inter alia* interactions between the 2'-OH of C10 and the H6 of U11, 2'OH of C3 and H8 of A4, and 2'-OH of G5 and H6 of C6. Additional cross-peaks are observed, but their unambiguous assignment is still challenging in this 2D NOESY spectrum due to a resonance overlap persisting even at 1 GHz. The incorporation of the HMT NOESY segment into a 3D experiment incorporating further information regarding the ribose protons, could potentially lift such ambiguities and contribute substantially to 3D structure determinations.

**Conclusions**

Hadamard MT was introduced here as an extension of previous projective-measurement experiments where, instead of a looped $t_1$ time-domain encoding, selective irradiations are used to impart significant sensitivity gains per unit-time in EXSY, NOESY and TOCSY homonuclear correlation experiments involving fast-exchanging protons. For achieving these gains HMT exploits the flow of fresh water polarization resetting the states of the targeted sites, whose perturbation away from equilibrium via saturations or inversions could then be used to magnify polarization transfer processes spreading throughout the molecules via dipole-dipole relaxation or through *J*-coupling. Thus, while chemical exchange with the solvent deteriorates conventional homonuclear transfer experiments, the abundant, slowly-relaxing water resonance improves these processes by several-fold when switching to this new encoding scheme. Relaxation properties of the non-labile protons on the receiving end of these transfers will limit the extent of these gains, as the MT process will only be effective over their 'memory times'. It follows that operating at ultrahigh fields, where $T_1$ relaxation times are usually longer, will improve the efficiency of these magnetization transfers. Further improvements at higher fields result from the line narrowing (in ppm) that labile $^1$Hs experience as a result of exchanges with the water, and from the possibility of increasing the extent of the saturation-derived transfer. Possible drawbacks of relying on such high-field MT processes concern potential sample heating effects, and an enhanced spin-diffusion among the non-labile sites in systems with high proton density. The former was not found to be a problem even when operating at 1 GHz; the latter, however, could rob the MT experiments from certain specificity. We investigated a similar possibility for the case of L-PROSY, where it was found that this new method still provided a 1:1 correlation with conventional NOESY cross-peak intensities when concerning translation of cross-peak intensities into inter-nuclear spatial distances –even in the presence of a spin-diffusion sink pool.[18] We assume that this correspondence will persist over a wide range of solvent exchange rates and correlation times, a matter that is still under investigation. It also remains to be investigated whether the continuous-irradiation or the repeated inversion versions of HMT, might perform differently in terms of their potential spin diffusion effects.

While this study focused on homonuclear transfers originating from labile $^1$Hs being replenished by the solvent, the HMT concept could be exploited in additional NMR settings where fast-relaxing sites can be individually addressed. These include methyl groups in otherwise deuterated proteins, inter-molecular interactions including protein- and drug-binding processes,[60–62] as well as fast-relaxing sites in paramagnetic biomolecules. The high efficiency and short-acquisition of HMT experiments can be further utilized for fast multidimensional reaction monitoring, especially involving small reagents/products in the fast tumbling regime that are traditionally hard to correlate with NOESY. The 2D Hadamard concepts introduced here could also be included as part of correlations with heteronuclei in 3D spectral acquisitions.[34,63] Heteronuclear analogues of the homonuclear polarization transfer processes discussed here can also be envisaged. Further discussions on all these cases will be presented in upcoming studies.

**Experimental section**

*Sample preparation.* Myo-inositol was purchased from Sigma Aldrich (Israel) and prepared as 5 mM solution at pH 6.0. 5 mM sucrose was prepared using household sugar at pH 6.5. Natural abundance α2-8 (SiA)$_4$ was purchased from EY Laboratories Inc (San Mateo, CA); 25 mg of this tetramer were dissolved in 400 μL of 20 mM phosphate buffer at pH 6.5 containing 0.05 % NaN$_3$, yielding an ~50 mM final solution at pH 7.35. The $^{13}$C/$^{15}$N labeled RNA encoding the 14mer gCUUGc tetraloop (5'-pppGGCAGCUUGCUGCC-3') was prepared from a linearized plasmid DNA by a run-off *in vitro* transcription using the T7 RNA polymerase.[64] In addition, the plasmid DNA contained a self-cleaving HDV ribozyme to ensure 3' homogeneity.[65] Labeled rNTPs were purchased from Silantes (Munich, Germany). The RNA was folded in NMR buffer (10 mM phosphate buffer + 1 mM EDTA pH: 6.4) in 90% H$_2$O and 10% D$_2$O by denaturing it for 5 min at 95°C and subsequently slowly cooling down to room temperature. The final concentration of the RNA was 1 mM. Ubiquitin was purchased from Asla Biotech and was dissolved in PBS (Dulbecco's Phosphate Buffer Saline at physiological pH, purchased from Biological Industries) at a concentration of 1 mM; LA5, the ligand binding domain 5 of the low-density lipoprotein receptor LDLR, was prepared as described by Szekely et al[55] at pH 7.4 and concentration 3 mM in 10 mM Tris buffer with 1 mM CaCl$_2$. All protein samples were prepared in H$_2$O/D$_2$O (90%:10%) solutions containing NaN$_3$.

*NMR experiments.* NMR experiments were conducted using either a 14.1 T Bruker magnet equipped with an Avance III console and TCI Prodigy probe; and on a 1GHz, 23.5 T Bruker Avance Neo equipped with a TCI cryoprobe. Hadamard experiments were carried out using H$_8$-H$_{64}$ Hadamard encoding matrices depending on the number of peaks in the spectrum. 10 Hz nutation field was used for saturation, while 20-25 Hz bandwidth *sinc1* inversion pulses were used in looped inversion method. Number of loops and duration saturation were determined according to T$_1$ values of receiving protons. Optimal values for NOESY and TOCSY mixing times were used in conventional experiments. Conventional TOCSY experiments were acquired using dipsi2gpph19/dipsi2esgpph standard Bruker sequences employing DIPSI2 isotropic mixing, while for NOESY experiments noesyfgpph19/noesyesgpph was used. In all cases, optimal mixing times and optimal WATERGATE delays for binomial water suppression according to the magnetic field were used. All spectra were processed in Bruker® TopSpin® 4.0.6. All spectra were apodized with QSINE or SINE window functions and while conventional spectra were zero-filled once, all Hadamard spectra were zero-filled to 256-1024 points.

**Acknowledgment:** We are grateful to Prof. D. Fass for the LA5 sample and to Dr. Tali Scherf (Weizmann Institute) for assistance in the GHz experiments. This work was supported by the Kimmel Institute for Magnetic Resonance (Weizmann Institute), the EU Horizon 2020 program (FET-OPEN Grant 828946, PATHOS), Israel Science Foundation Grant 965/18, and the Perlman Family Foundation. HS was supported by DFG-funded collaborative research center 902. Work at BMRZ is supported by the state of Hesse. Joint support to LF, HS was given by the German-Israel Foundation (grant G-1501-302). We wish to thank Boris Fürtig, Robbin Schnieders and Christian Richter for stimulating discussions.

**Supporting Information**

In general, HMT enhancements will be determined by the chemical exchange rate, and absolute values of the self- and cross-relaxation rates. Small relaxation rates and fast rates of chemical exchange will significantly impact the efficiency of NOESY experiments, yet these are the scenarios where one can exploit the fullest potential of Anti-Zeno Effects and MT encoded measurements as shown with myo-inositol in Figure 3. Figure S1 illustrates the achievable enhancements that HMT can provide with respect to various rotational correlation times and chemical exchange rates, when normalized against conventional NOEs. Figure S1a shows these enhancements assuming a continuous saturating RF with nutation field $\omega_1 = 20 \cdot 2\pi \, rad/s$. One can see that the enhancements show a non-monotonic behavior with respect to correlation times, which reflects the similar non-monotonic dependencies exhibited by the cross-relaxation rates with $\tau_c$ (Fig S1a, inset). Figure S1b show a similar plot but for the case when Hadamard MT is achieved by looped inversion schemes, confirming similar maximum enhancements and overall behavior. Notice that as solvent exchange rates become faster both approaches increase their maximal potential enhancements, eventually surpassing ≥10-fold magnifications; this explains the strong cross-peak intensities achieved in the experiments, in only a fraction of the time needed for a conventional acquisition. Since these substantial enhancements are maximized when cross-relaxation rates are weak, it can be anticipated that they will not only arise from short correlation times (where σ become smaller) but also for large internuclear distances making the dipole-dipole coupling constant progressively weaker.

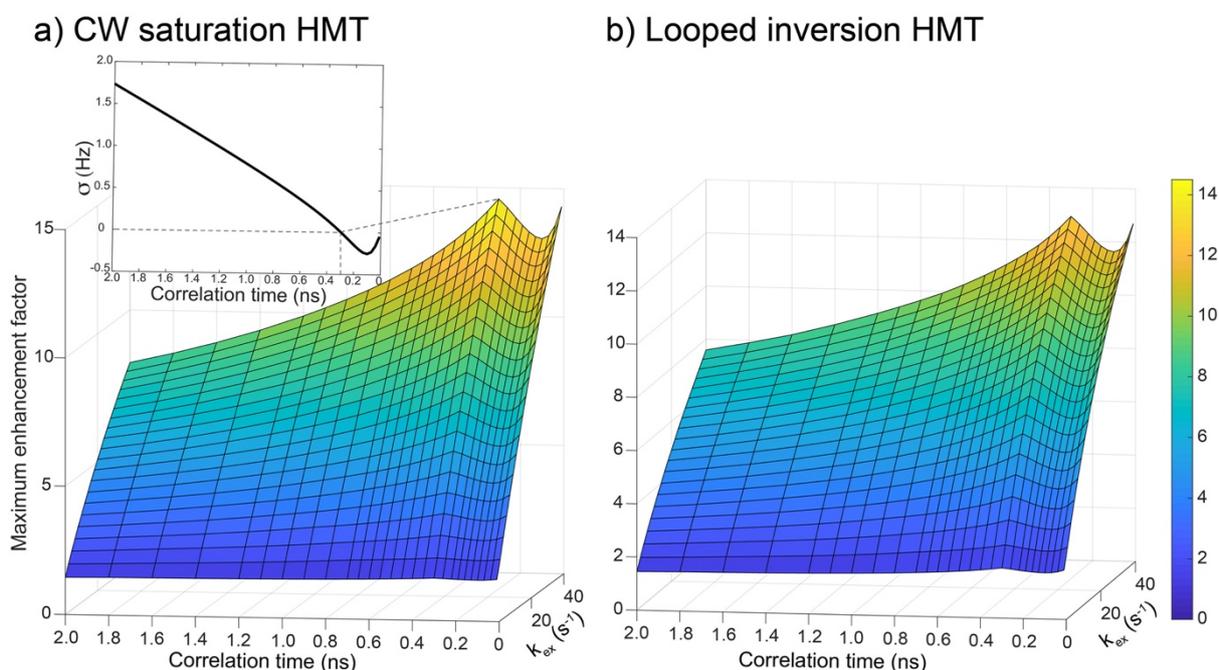

**Figure S1.** a) Maximum achievable enhancements of Hadamard MT NOESY with respect to correlation times and chemical exchange rates. The non-monotonic change shown by the enhancement with respect to correlation times can be explained by the non-monotonic cross-relaxation rates' dependence on the $\tau_c$, as shown in the inset. b) Similar enhancements plots but upon using looped inversions for the HMT instead of a CW saturation pulse. Parameters were similar as in Figure 2: cross-relaxation rates were calculated for internuclear distance of $r = 2$ Å at 14.1 T. Relaxation rates were chosen to be $T_{2_l} = T_{2_{nl}} = 0.3 \, s$, $T_{1_l} = 0.5 \, s$ and $T_{1_{nl}} = 0.8 \, s$, while water relaxation constants were taken to be $T_{2_w} = 0.5 \, s$ and $T_{1_w} = 3 \, s$.

This provides HMT with an opportunity to reveal previously undetectable cross-peaks in challenging systems, as evidenced by the various experiments in the main text.

Further insight onto these joint correlation time / internuclear distances effects are presented in Figure S2, calculated under the assumption of a continuous-wave saturation. The magnitude of the normalized transferred magnetization is shown in Figure S2a, plotted vs correlation times and internuclear distances for a water chemical exchange rate of 40 s$^{-1}$. As expected, the strongest cross-peaks can then be detected for slow tumbling times and short internuclear distances that make the cross-relaxation rates the strongest. By contrast, the HMT *enhancements* vs a conventional measurement (Fig. S2b) show exactly the opposite scenario: they are the smallest for strongest cross-relaxation rates and vice-versa. For this instance, HMT enhancements are >14-fold regardless of the means of encoding; thus, a conventional experiment with similar sensitivity would take almost 200 times longer to acquire. Moreover, if one arbitrarily defines 0.5% of the total magnetization intensity as an arbitrary experimental detection threshold, Figure S2a provides an estimate of the range of distances that HMT could target. Assuming a medium-sized biomolecule with correlation time ~3 ns this cross-peak intensity threshold would provide correlations with the labile $^1$H of up to 5.2 Å; in a fast tumbling regime where correlation times are ~0.1 ns, this distance drops to 3.5 Å. These figures are to be compared against what similar thresholds and models predict for a conventional NOESY experiment: 3.6 and 2.5 Å, respectively for each particular case.

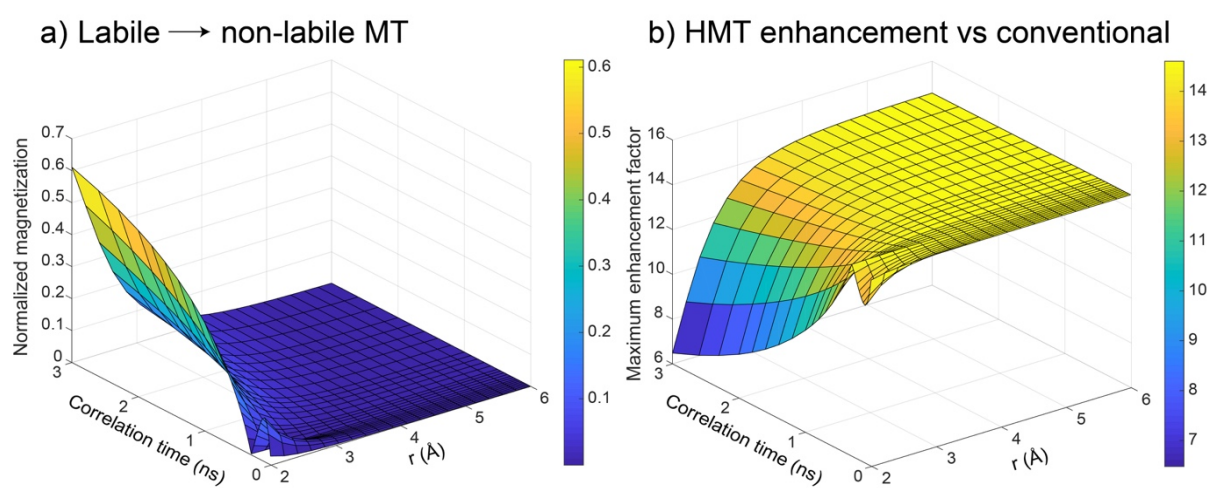

**Figure S2.** a) Normalized magnetization transferred by continuous RF saturation from labile to non-labile protons plotted with respect to correlation times and internuclear distances. b) When these magnetization values are normalized to conventional NOESY, expected Hadamard MT saturation enhancements are obtained showing anticorrelation with the surface plot in a). A fast exchange rate of 40 s$^{-1}$ was used in these simulations.

Equations 4, derived to estimate the efficiency of looped inversion MT experiments, can also be used to estimate the enhancements achievable in HMT-based versions of the TOCSY experiment. Indeed, as TOCSY requires a constant spin-lock to enable the transfer through *J*-couplings, it follows that only such looped inversions (followed by isotropic mixing sequences) are compatible with these schemes. To estimate the resulting enhancements, we replaced the cross-relaxation rate connecting non-labile and labile proton pools in Eqs. (4) with an average *J*-coupling-based transfer rate, and the $T_1$ values appearing in them with the spin-lattice relaxation times in the rotating frame ($T_{1\rho}$). Figure S3 shows the expected enhancements in HMT versions of the TOCSY experiment, for two different coupling transfer rates: 7 Hz, representing a typical three-bond coupling values, and 2 Hz, representing a longer-range proton-proton coupling. Notice that these HMT TOCSY experiments end up achieving somewhat smaller enhancements than their NOESY counterparts; this is due to the faster relaxation process (i.e., $T_{1\rho} < T_1$), constraining the extent of the transfer. Notice as well that higher enhancements are obtained for smaller values in the J-coupling (Figure S3b), reflecting

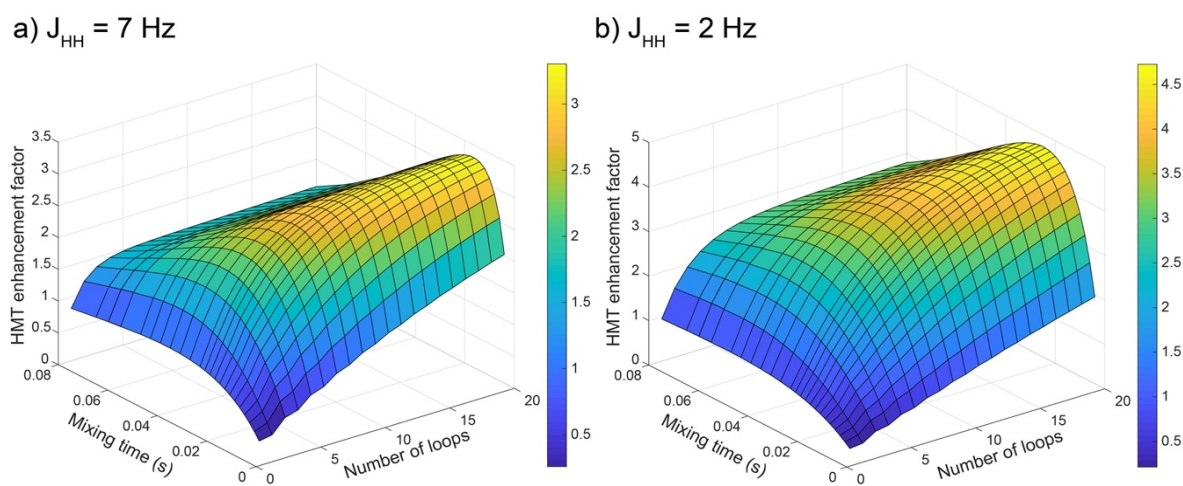

**Figure S3.** Enhancements achievable in Hadamard MT TOCSY experiments comparing to conventional experiments for water exchange rate of 40 s$^{-1}$ and two different *J*-coupling rates: a) 7 Hz which is typical $^3J$ coupling constant and b) 2 Hz representing long range coupling. Relaxation rates were $T_{1\rho_l} = T_{1\rho_{nl}} = 0.3\ s$, and $T_{1\rho_w} = 0.5\ s$

the fact that the chemical exchange involved will average out these weaker couplings, and hence further reduce SNR in conventional TOCSY transfers than in their looped counterparts.

As complement to Figure 4 in the main text, Figure S4 compares conventional and HMT TOCSY experiments recorded on (SiA)$_4$. Notice the superior quality and information content provided by the HMT experiment despite its ≥30-fold shorter acquisition time than that of the conventional experiment. In this case the HMT TOCSY yields correlations with labile protons that are up to 5 bonds away (i.e. IOH7-I9), despite the presence of >100 s$^{-1}$ chemical exchanges with water.

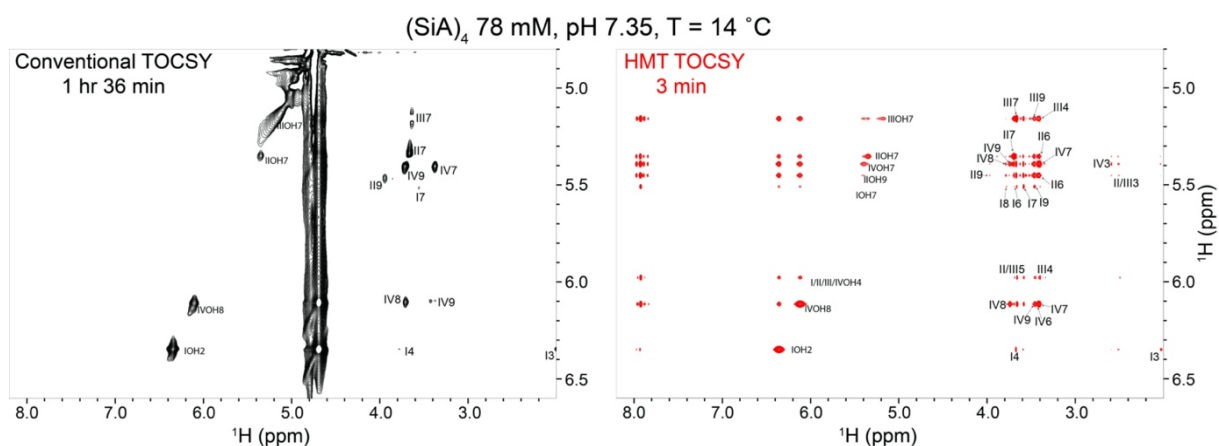

Figure S4. TOCSY spectra acquired on 78 mM (SiA)$_4$ using conventional (40 ms DIPSI2) and Hadamard MT TOCSY (16x12 ms DIPSI2). While chemical exchange averages out *J*-couplings rendering TOCSY experiments very inefficient, HMT TOCSY provided numerous TOCSY cross-peaks that help in both peak assignments and the primary structure determinations. Spectra were acquired at 1GHz on an Avance NEO equipped with a room temperature TCI probe.

Figure S5 further illustrates the advantages of performing HMT experiments at high magnetic fields, when targeting nucleic acids imino resonances like those in the 14mer hairpin RNA studied in Figure 5 (main text). While at 14.1 T peaks such as those of G9H1 and G2H1 are overlapping, they are fully resolved at 23.5 T, enabling their HMT encoding. Furthermore, the SNR of these and other imino peaks at 1 GHz is higher than at 600 MHz. Figures S5b and S5c compare conventional imino NOESY spectra against HMT-based counterparts at these two fields. While HMT is superior at both magnetic fields, it can be appreciated from the extracted 1D projections that its enhancements are ca. 50% larger at 1 GHz than at 600 MHz – comparing in both cases against conventional experiments done in the same field. *This reflects the multiple factors mentioned in the main text, concerning the supra-linear improvements that HMT achieves with magnetic field.*

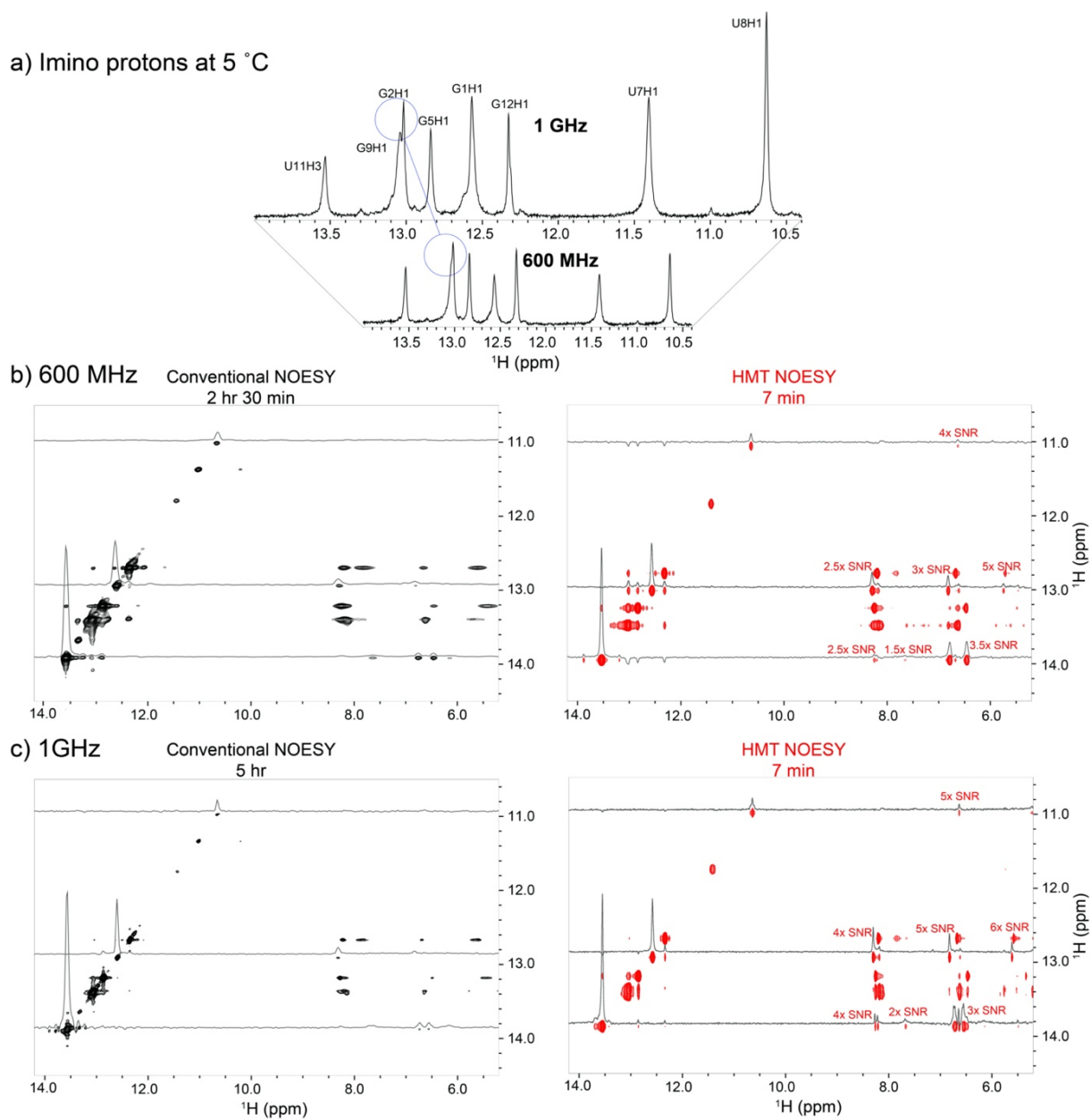

Figure S5. Magnetic field benefits for nucleic acid experiments, as illustrated on the imino resonances of a 14mer RNA. a) Resolution improvement provided by 1 GHz over 600 MHz; highlighting the resolution of G9H1 and G2H1 and enabling their distinction in HMT correlations. b) Conventional and HMT NOESY spectra acquired at 600 MHz and with 100 ms mixing, the latter using a 15 loop of inversion pulses with 12 Hz bandwidth. c) Idem for 1 GHz, using 14 loops and 20 Hz inversion pulses (allowed thanks to the better peak separation).

Complementing the HMT data in Figures 7a and 8b, Figure S6a shows ubiquitin's WATERGATE 3919[40–42] spectrum highlighting the 6-10 ppm $^1$H region, and the same spectrum after suppression of all $^{15}$N/$^{13}$C bound protons so as to reveal potential hydroxyl signatures. After examining these peaks at different temperatures (Fig S6b), it is possible to tentatively assign as hydroxyl protons those peaks that become sharper at lower temperatures; all the amide, amino and aromatic peaks are suppressed by multiple quantum filter, while the peaks that remain unchanged with temperature are attributed to unlabeled impurities present in the sample. Figure S6c shows NOESY spectrum from Figure 6b in the main text zoomed on amino protons region showing exchange cross-peaks with another set of amines and some correlations to aliphatic proton region. Figure S6d similarly shows hydroxyl protons region of conventional NOESY spectrum but acquired after suppression of $^{15}$N/$^{13}$C-bound protons compared to HMT spectrum in Figure 7b.

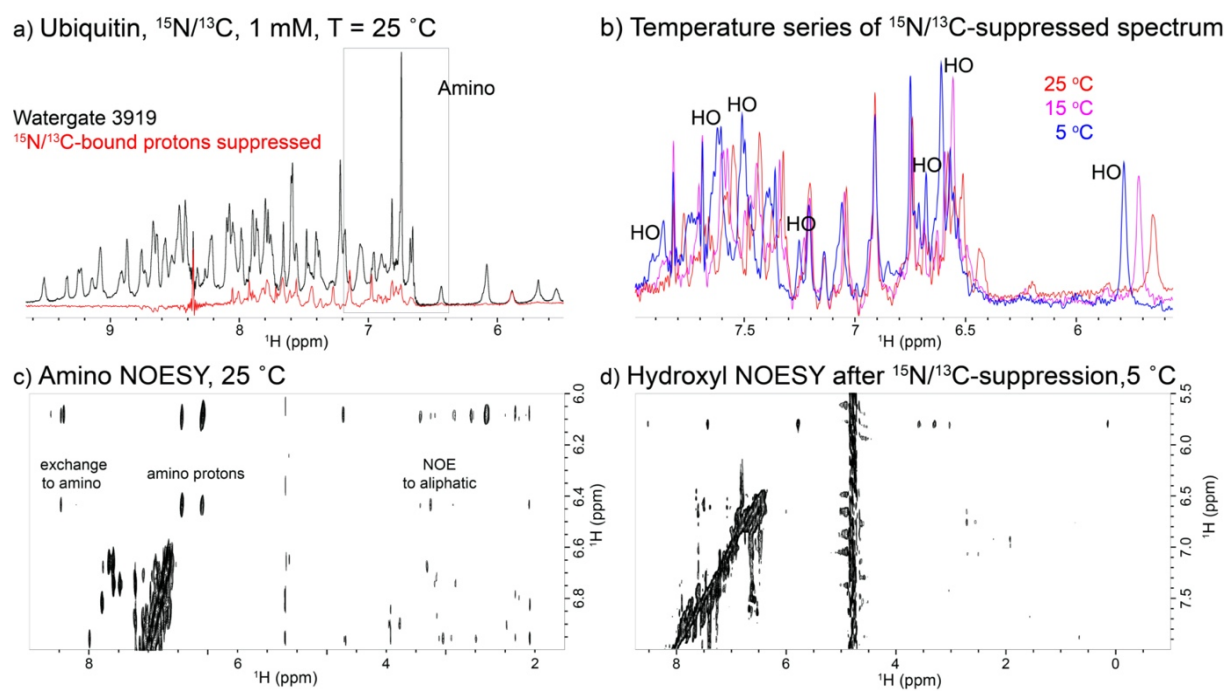

Figure S6. a) Part of a $^{15}$N/$^{13}$C decoupled WATERGATE 3919 spectrum of Ubiquitin illustrating amide and amine protons (black) at 25 °C and what remains from it after suppression of $^{15}$N/$^{13}$C-bound protons. Shown by the dashed rectangle is the amino resonances region. b) Temperature series of corresponding suppressed spectrum revealing peaks that are getting sharper at lower temperatures as potential hydroxyl protons. c) Part of NOESY spectrum from Figure 6b zoomed on amino region. d) NOESY spectrum acquired after suppression of $^{15}$N/$^{13}$C-bound protons using mixing time of 150 ms with total acquisition time of 4 hours.

Figure S7 shows conventional full NOESY spectrum of 14mer hairpin RNA studied in the main text and two regions concentrating on regions where amino (Fig S7b) and hydroxyl (Fig S7c) protons resonate. While amino protons show cross-peaks to imino and other pair of amino protons, correlations involving hydroxyl protons is harder to discern because they overlap with amino resonances. The intrinsically selective nature of Hadamard encoding helps target specific protons, and in combination with multiple quantum filter for $^{15}$N/$^{13}$C-bound proton suppression, it can yield correlations involving solely hydroxyl or amino protons.

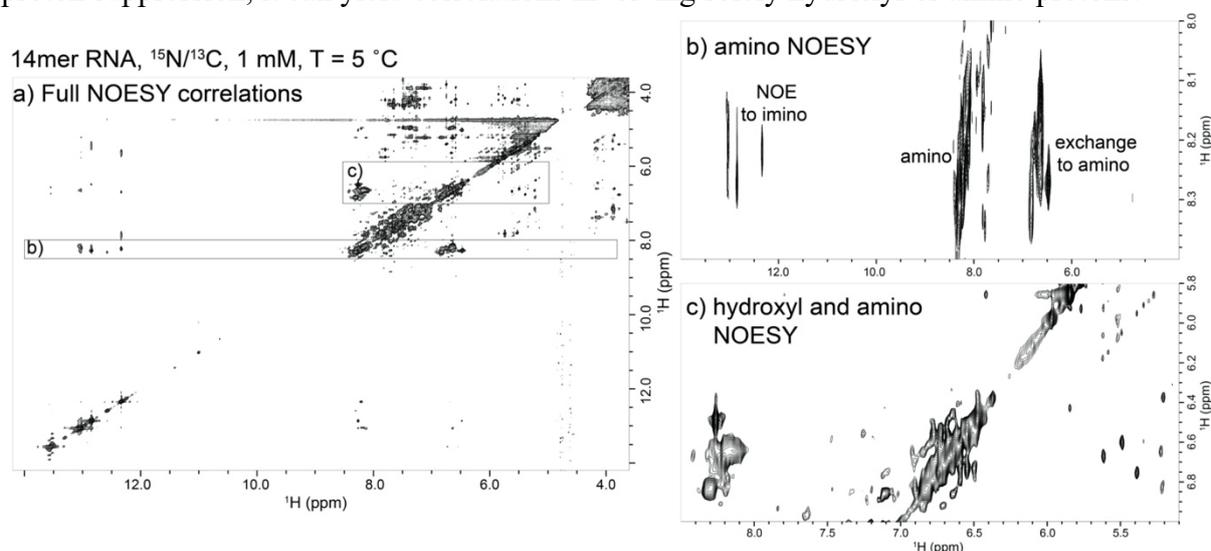

Figure S7. a) Full NOESY spectrum acquired on 14mer hairpin RNA using 100 ms mixing and 90 us delay for a binomial water suppression. b) Zoomed amino region of NOESY spectrum illustrating exchange correlations to amino protons and NOE cross-peaks to imino protons. c) Portion of NOESY spectrum containing hydroxyl protons that overlap with amino resonances. Notice how difficult it is to discern these peaks from amino correlations in conventional spectra, when there is no suppression of $^{15}$N/$^{13}$C-bound protons.

**References**
(1) Jeener, J.; Meier, B. H.; Bachmann, P.; Ernst, R. R. Investigation of Exchange Processes by Two-Dimensional NMR Spectroscopy. *Journal of Chemical Physics* **1979**, *71* (11), 4546.
(2) Aue, W. P.; Bartholdi, E.; Ernst, R. R. Two-Dimensional Spectroscopy. Application to Nuclear Magnetic Resonance. *The Journal of Chemical Physics* **1976**, *64* (5), 2229.
(3) Macura, S.; Huang, Y.; Suter, D.; Ernst, R. R. Two-Dimensional Chemical Exchange and Cross-Relaxation Spectroscopy of Coupled Nuclear Spins. *Journal of Magnetic Resonance (1969)* **1981**, *43* (2), 259.
(4) Macura, S.; Ernst, R. R. Elucidation of Cross Relaxation in Liquids by Two-Dimensional N.M.R. Spectroscopy. *Molecular Physics* **1980**, *41* (1), 95.
(5) Overhauser, A. W. Polarization of Nuclei in Metals. *Physical Review* **1953**, *92* (2), 411.
(6) Noggle, J. H.; Schirmer, R. E. *The Nuclear Overhauser Effect*; ACADEMIC PRESS INC., 1971.
(7) Bax, A.; Davis, D. G. Practical Aspects of Two-Dimensional Transverse NOE Spectroscopy. *Journal of Magnetic Resonance (1969)* **1985**, *63* (1), 207.
(8) Neuhaus, D.; Williamson, M. P. *The Nuclear Overhauser Effect in Structural and Conformational Analysis*; 1989; Vol. 75.
(9) Cavanagh, J.; Fairbrother, W. J.; Palmer, A. G.; Rance, M.; J., S. N. *Protein NMR Spectroscopy*; Elsevier, 2007.
(10) Wuthrich, K. *NMR of Proteins and Nucleic Acids*; Wiley, 1986.
(11) Kumar, A.; Wagner, G.; Ernst, R. R.; Wuethrich, K. Buildup Rates of the Nuclear Overhauser Effect Measured by Two-Dimensional Proton Magnetic Resonance Spectroscopy: Implications for Studies of Protein Conformation. *Journal of the*

*American Chemical Society* **1981**, *103* (13), 3654.

(12) van de Ven, F. J. M.; Blommers, M. J. J.; Schouten, R. E.; Hilbers, C. W. Calculation of Interproton Distances from NOE Intensities. A Relaxation Matrix Approach without Requirement of a Molecular Model. *Journal of Magnetic Resonance (1969)* **1991**, *94* (1), 140.

(13) Huang, Y.; Macura, S.; Ernst, R. R. Carbon-13 Exchange Maps for the Elucidation of Chemical Exchange Networks. *Journal of the American Chemical Society* **1981**, *103* (18), 5327.

(14) Macura, S.; Westler, W. M.; Markley, J. L. Two-Dimensional Exchange Spectroscopy of Proteins. *Methods in Enzymology* **1994**, *239* (C), 106.

(15) Meier, B. H.; Ernst, R. R. Elucidation of Chemical Exchange Networks by Two-Dimensional NMR Spectroscopy: The Heptamethylbenzenonium Ion. *Journal of the American Chemical Society* **1979**, *101* (21), 6441.

(16) Rucker, S.; Shaka, A. J. Broadband Homonuclear Cross Polarization in 2D N.M.R. Using DIPSI-2. *Molecular Physics* **1989**, *68* (2), 509.

(17) Furrer, J.; Kramer, F.; Marino, J. P.; Glaser, S. J.; Luy, B. Homonuclear Hartmann-Hahn Transfer with Reduced Relaxation Losses by Use of the MOCCA-XY16 Multiple Pulse Sequence. *Journal of Magnetic Resonance* **2004**, *166* (1), 39.

(18) Novakovic, M.; Cousin, S. F.; Jaroszewicz, M. J.; Rosenzweig, R.; Frydman, L. Looped-PROjected SpectroscopY (L-PROSY): A Simple Approach to Enhance Backbone/Sidechain Cross-Peaks in 1H NMR. *Journal of Magnetic Resonance* **2018**, *294*, 169.

(19) Milburn, G. J. Quantum Zeno Effect and Motional Narrowing in a Two-Level System. *Journal of the Optical Society of America B* **1988**, *5* (6), 1317.

(20) Itano, W. M.; Heinzen, D. J.; Bollinger, J. J.; Wineland, D. J. Quantum Zeno Effect. *Physical Review A* **1990**, *41* (5), 2295.

(21) Facchi, P.; Lidar, D. A.; Pascazio, S. Unification of Dynamical Decoupling and the Quantum Zeno Effect. *Physical Review A* **2004**, *69* (3), 032314.

(22) Facchi, P.; Pascazio, S. Quantum Zeno Dynamics: Mathematical and Physical Aspects. *Journal of Physics A: Mathematical and Theoretical* **2008**, *41* (49).

(23) Zheng, W.; Xu, D. Z.; Peng, X.; Zhou, X.; Du, J.; Sun, C. P. Experimental Demonstration of the Quantum Zeno Effect in NMR with Entanglement-Based Measurements. *Physical Review A - Atomic, Molecular, and Optical Physics* **2013**, *87* (3), 1.

(24) Álvarez, G. A.; Rao, D. D. B.; Frydman, L.; Kurizki, G. Zeno and Anti-Zeno Polarization Control of Spin Ensembles by Induced Dephasing. *Physical Review Letters* **2010**, *105* (16), 1.

(25) Bretschneider, C. O.; Alvarez, G. A.; Kurizki, G.; Frydman, L. Controlling Spin-Spin Network Dynamics by Repeated Projective Measurements. *Physical Review Letters* **2012**, *108* (14), 1.

(26) Schanda, P.; Brutscher, B. Very Fast Two-Dimensional NMR Spectroscopy for Real-Time Investigation of Dynamic Events in Proteins on the Time Scale of Seconds. *Journal of the American Chemical Society* **2005**, *127*, 8014.

(27) Schanda, P.; Kupče, Ē.; Brutscher, B. SOFAST-HMQC Experiments for Recording Two-Dimensional Deteronuclear Correlation Spectra of Proteins within a Few Seconds. *Journal of Biomolecular NMR* **2005**, *33* (4), 199.

(28) Kupče, E.; Freeman, R. Fast Multidimensional NMR by Polarization Sharing. *Magnetic Resonance in Chemistry* **2007**, *45* (1), 2.

(29) Schulze-Sünninghausen, D.; Becker, J.; Luy, B. Rapid Heteronuclear Single Quantum Correlation NMR Spectra at Natural Abundance. *Journal of the American Chemical*

*Society* **2014**, *136* (4), 1242.
(30) Guivel-Scharen, V.; Sinnwell, T.; Wolff, S. D.; Balaban, R. S. Detection of Proton Chemical Exchange between Metabolites and Water in Biological Tissues. *Journal of magnetic resonance (San Diego, Calif. : 1997)* **1998**, *133*, 36.
(31) Zhou, J.; Zijl, P. C. M. van. Chemical Exchange Saturation Transfer Imaging and Spectroscopy. *Progress in Nuclear Magnetic Resonance Spectroscopy* **2006**, *48* (2–3), 109.
(32) Martinho, R. P.; Novakovic, M.; Olsen, G. L.; Frydman, L. Heteronuclear 1D and 2D NMR Resonances Detected by Chemical Exchange Saturation Transfer to Water. *Angewandte Chemie International Edition* **2017**, *56* (13), 3521.
(33) Novakovic, M.; Martinho, R. P.; Olsen, G. L.; Lustig, M. S.; Frydman, L. Sensitivity-Enhanced Detection of Non-Labile Proton and Carbon NMR Spectra on Water Resonances. *Phys. Chem. Chem. Phys.* **2017**.
(34) Kupče, E.; Nishida, T.; Freeman, R. Hadamard NMR Spectroscopy. *Progress in Nuclear Magnetic Resonance Spectroscopy* **2003**, *42* (3–4), 95.
(35) Kupče, Ē.; Freeman, R. Two-Dimensional Hadamard Spectroscopy. *Journal of Magnetic Resonance* **2003**, *162* (1), 300.
(36) Kupče, E.; Freeman, R. Frequency-Domain Hadamard Spectroscopy. *Journal of Magnetic Resonance* **2003**, *162* (1), 158.
(37) Zhou, Z.; Kümmerle, R.; Qiu, X.; Redwine, D.; Cong, R.; Taha, A.; Baugh, D.; Winniford, B. A New Decoupling Method for Accurate Quantification of Polyethylene Copolymer Composition and Triad Sequence Distribution with 13C NMR. *Journal of Magnetic Resonance* **2007**, *187* (2), 225.
(38) Yuwen, T.; Skrynnikov, N. R. CP-HISQC: A Better Version of HSQC Experiment for Intrinsically Disordered Proteins under Physiological Conditions. *Journal of Biomolecular NMR* **2014**, *58*, 175.
(39) Hwang, T.; Shaka, A. J. Water Suppression That Works. Excitation Sculpting Using Arbitrary Wave-Forms and Pulsed-Field Gradients. *Journal of Magnetic Resonance - Series A* **1995**, *112* (2).
(40) Davis, A. L.; Wimperis, S. A Solvent Suppression Technique Giving NMR Spectra with Minimal Amplitude and Phase Distortion. *Journal of Magnetic Resonance (1969)* **1989**, *84* (3), 620.
(41) Piotto, M.; Saudek, V.; Sklenář, V. Gradient-Tailored Excitation for Single-Quantum NMR Spectroscopy of Aqueous Solutions. *Journal of Biomolecular NMR* **1992**, *2* (6), 661.
(42) Lippens, G.; Dhalluin, C.; Wieruszeski, J. M. Use of a Water Flip-Back Pulse in the Homonuclear NOESY Experiment. *Journal of Biomolecular NMR* **1995**, *5* (3), 327.
(43) McConnell, H. M. Reaction Rates by Nuclear Magnetic Resonance. *The Journal of Chemical Physics* **1958**, *28* (3), 430.
(44) Helgstrand, M.; Hard, T.; Allard, P. Simulations of NMR Pulse Sequences during Equilibrium and Non-Equilibrium Chemical Exchange. *Journal of Biomolecular NMR* **2000**, *18*, 49.
(45) Zaiss, M.; Bachert, P. Chemical Exchange Saturation Transfer (CEST) and MR Z-Spectroscopy in Vivo: A Review of Theoretical Approaches and Methods. *Phys. Med. Biol* **2013**, *58*, 221.
(46) Novakovic, M.; Martinho, R. P.; Olsen, G. L.; Lustig, M. S.; Frydman, L. Sensitivity-Enhanced Detection of Non-Labile Proton and Carbon NMR Spectra on Water Resonances. *Physical Chemistry Chemical Physics* **2018**, *20* (1), 56.
(47) Levitt, M. . *Spin Dynamics: Basics of Nuclear Magnetic Resonance*; 2000.
(48) Roussel, T.; Rosenberg, J. T.; Grant, S. C.; Frydman, L. Brain Investigations of Rodent

Disease Models by Chemical Exchange Saturation Transfer at 21.1 T. *NMR in Biomedicine* **2018**, *31* (11), e3995.

(49) Chung, J. J.; Choi, W.; Jin, T.; Lee, J. H.; Kim, S.-G. Chemical-Exchange-Sensitive MRI of Amide, Amine and NOE at 9.4 T versus 15.2 T. *NMR in Biomedicine* **2017**, *30* (9), e3740.

(50) van Zijl, P. C. M.; Lam, W. W.; Xu, J.; Knutsson, L.; Stanisz, G. J. Magnetization Transfer Contrast and Chemical Exchange Saturation Transfer MRI. Features and Analysis of the Field-Dependent Saturation Spectrum. *NeuroImage* **2018**, *168* (April 2017), 222.

(51) Korb, J. P.; Bryant, R. G. Magnetic Field Dependence of Proton Spin-Lattice Relaxation Times. *Magnetic Resonance in Medicine* **2002**, *48* (1), 21.

(52) Kowalewski, J.; Maler, L. *Nuclear Spin Relaxation in Liquids*; CRC Press: Cambridge, 2006.

(53) Battistel, M. D.; Shangold, M.; Trinh, L.; Shiloach, J.; Freedberg, I. Evidence for Helical Structure in a Tetramer of α 2-8 Sialic Acid: Unveiling a Structural Antigen. *Journal of the American Chemical Society* **2012**, *134*, 8.

(54) Shinar, H.; Battistel, M. D.; Mandler, M.; Lichaa, F.; Freedberg, D. I.; Navon, G. Sialo-CEST: Chemical Exchange Saturation Transfer NMR of Oligo- and Poly-Sialic Acids and the Assignment of Their Hydroxyl Groups Using Selective- and HSQC-TOCSY. *Carbohydrate Research* **2014**, *389* (1), 165.

(55) Szekely, O.; Armony, G.; Olsen, G. L.; Bigman, L. S.; Levy, Y.; Fass, D.; Frydman, L. Identification and Rationalization of Kinetic Folding Intermediates for a Low-Density Lipoprotein Receptor Ligand-Binding Module. *Biochemistry* **2018**, *57* (32), 4776.

(56) Hennig, M.; Fohrer, J.; Carlomagno, T. Assignment and NOE Analysis of 2′-Hydroxyl Protons in RNA : Implications for Stabilization of RNA A-Form Duplexes. *Journal of American Chemical Society* **2005**, 2028.

(57) Grzesiek, S.; Bax, A. Measurement of Amide Proton Exchange Rates and NOEs with Water in 13C/15N-Enriched Calcineurin B. *Journal of Biomolecular NMR* **1993**, *3* (6), 627.

(58) Wang, Y. X.; Freedberg, D. I.; Grzesiek, S.; Torchia, D. A.; Wingfield, P. T.; Kaufman, J. D.; Stahl, S. J.; Chang, C. H.; Hodge, C. N. Mapping Hydration Water Molecules in the HIV-1 Protease/DMP323 Complex in Solution by NMR Spectroscopy. *Biochemistry* **1996**, *35* (39), 12694.

(59) Nozinovic, S.; Fürtig, B.; Jonker, H. R. A.; Richter, C.; Schwalbe, H. High-Resolution NMR Structure of an RNA Model System: The 14-Mer CUUCGg Tetraloop Hairpin RNA. *Nucleic Acids Research* **2009**, *38* (2), 683.

(60) Dalvit, C.; Pevarello, P.; Tato, M.; Veronesi, M.; Vulpetti, A.; Sundström, M. Identification of Compounds with Binding Affinity to Proteins via Magnetization Transfer from Bulk Water. *Journal of Biomolecular NMR* **2000**, *18* (1), 65.

(61) Dalvit, C.; Fogliatto, G.; Stewart, A.; Veronesi, M.; Stockman, B. WaterLOGSY as a Method for Primary NMR Screening: Practical Aspects and Range of Applicability. *Journal of Biomolecular NMR* **2001**, *21* (4), 349.

(62) Mayer, M.; Meyer, B. Characterization of Ligand Binding by Saturation Transfer Difference NMR Spectroscopy. *Angewandte Chemie - International Edition* **1999**, *38* (12), 1784.

(63) Brutscher, B. Combined Frequency- and Time-Domain NMR Spectroscopy. Application to Fast Protein Resonance Assignment. *Journal of Biomolecular NMR* **2004**, *29*, 57.

(64) Guilleres, J.; Lopez, P. J.; Proux, F.; Launay, H.; Dreyfus, M. A Mutation in T7 RNA Polymerase That Facilitates Promoter Clearance. *Proceedings of the National*

*Academy of Sciences of the United States of America* **2005**, *102* (17), 5958.

(65) Walker, Scott. C.; Avis, Johanna. M.; Conn, G. L. General Plasmids for Producing RNA in Vitro Transcripts with Homogeneous Ends. *Nucleic Acids Research* **2003**, *31* (15), 82e.